\documentclass{aa}

\usepackage{graphicx}
\usepackage{txfonts}
\usepackage{amsmath}
\usepackage{graphicx}
\usepackage{txfonts}
\usepackage{natbib, twoopt}
\usepackage{subfig}
\usepackage{color}
\usepackage{mathrsfs}
\usepackage{xspace}
\usepackage{rotating}
\usepackage{lscape}
\usepackage{amssymb}
\usepackage{longtable}
\usepackage{pifont}
\usepackage{multirow}
\usepackage{adjustbox}
\usepackage[switch]{lineno} % switch means 2 columns
 \newcommand*\patchAmsMathEnvironmentForLineno[1]{%
   \expandafter\let\csname old#1\expandafter\endcsname\csname #1\endcsname
   \expandafter\let\csname oldend#1\expandafter\endcsname\csname end#1\endcsname
   \renewenvironment{#1}%
      {\linenomath\csname old#1\endcsname}%
      {\csname oldend#1\endcsname\endlinenomath}}% 
 \newcommand*\patchBothAmsMathEnvironmentsForLineno[1]{%
   \patchAmsMathEnvironmentForLineno{#1}%
      \patchAmsMathEnvironmentForLineno{#1*}
 }%
 \AtBeginDocument{%
 \patchBothAmsMathEnvironmentsForLineno{equation}%
 \patchBothAmsMathEnvironmentsForLineno{align}%
 \patchBothAmsMathEnvironmentsForLineno{flalign}%
 \patchBothAmsMathEnvironmentsForLineno{alignat}%
 \patchBothAmsMathEnvironmentsForLineno{gather}%
 \patchBothAmsMathEnvironmentsForLineno{multline}%
}

\usepackage{gensymb}
\usepackage[plainpages=False,bookmarks=True,colorlinks=True,citecolor=blue,linkcolor=blue,urlcolor=magenta,breaklinks=True,pagebackref=False]{hyperref}
\usepackage[mathscr]{euscript}

\newcommand{\GitLinkTreegp}{\url{https://github.com/PFLeget/treegp}}

\newcommand{\numbervisitsSSP}{2294 }
\newcommand{\rmsastroHSC}{33 \xspace}
\newcommand{\magmagic}{20.5 }

\newcommand{\bs}[1]{\boldsymbol{#1}}

\defcitealias{Bernstein17}{B17}
\defcitealias{fortino2020reducing}{F20}

\begin{document}

 \graphicspath{{figures/}} 

 \title{Improving the astrometric solution of the Hyper Suprime-Cam with anisotropic Gaussian processes}
   \titlerunning{Astrometry \& Gaussian processes}
   \authorrunning{P.-F. L\'eget et al.}
   \date{Received 02 February 2021 / Accepted 17 March 2021}
\author{P.-F.~L\'eget \inst{\ref{lpnhe},\ref{kipac}}
\and   P.~Astier \inst{\ref{lpnhe}}
\and   N.~Regnault \inst{\ref{lpnhe}}
\and   M.~Jarvis \inst{\ref{upenn}}
\and   P.~Antilogus \inst{\ref{lpnhe}}
\and   A.~Roodman \inst{\ref{kipac}, \ref{slac}} 
\and   D.~Rubin \inst{\ref{Hawaii}, \ref{lbnl}}
\and   C.~Saunders \inst{\ref{princeton}, \ref{lpnhe}}
}

\institute{\tiny
LPNHE, CNRS/IN2P3, Sorbonne Universit\'e, Laboratoire de
Physique Nucl\'eaire et de Hautes \'Energies, F-75005, Paris, France \label{lpnhe}
\and
    Kavli Institute for Particle Astrophysics and Cosmology,
    Department of Physics, Stanford University, 
    Stanford, CA 94305 \label{kipac}
\and 
Department of Physics and Astronomy, University of Pennsylvania, Philadelphia, PA 19104, USA \label{upenn}
\and    
    Department of Astrophysical Sciences, Princeton University, 4 Ivy Lane, Princeton NJ 08544, USA \label{princeton}
\and 
SLAC National Accelerator Laboratory, Menlo Park, CA 94025, USA \label{slac}
\and
Department of Physics and Astronomy, University of Hawai‘i at Manoa, Honolulu, Hawai‘i 96822, USA \label{Hawaii}
\and
E.O. Lawrence Berkeley National Laboratory, 1 Cyclotron Rd., Berkeley, CA 94720, USA \label{lbnl}
}

\abstract
{We study astrometric residuals from a simultaneous fit of Hyper Suprime-Cam images.}
{We aim to characterize these residuals and study the extent to which they are
   dominated by atmospheric contributions for bright sources.}
{We use Gaussian process interpolation, with a correlation
  function (kernel), measured from the data,  to smooth and correct the observed  astrometric residual field.}
{We find that Gaussian process interpolation with a von K\'arm\'an kernel
  allows us to reduce the covariances of astrometric residuals for nearby sources by about
  one order of magnitude, from 30 mas$^2$ to 3 mas$^2$ at angular scales of $\sim$~1 arcmin, and to halve the r.m.s.~residuals. Those reductions using Gaussian process interpolation are similar to recent result published with the Dark Energy Survey dataset. We are then able to detect the small static  astrometric residuals due
  to the Hyper Suprime-Cam sensors effects. We discuss how the Gaussian process interpolation of astrometric residuals impacts galaxy shape measurements, in particular in the context of cosmic shear analyses at the Rubin Observatory Legacy Survey of Space and Time.}  {}  \maketitle

\section{Introduction}
\label{introduction}

Astrometry refers to the determination of the position of
astronomical sources on the sky. In imaging surveys, a
crucial step in astrometry is the determination of the mapping of
coordinates measured in pixel space on the sensors to a celestial
coordinate system. Measurements of source positions with the
sensors and determination of the mapping are both affected by
uncertainties that may have consequences on down-stream measurements
performed on the images, especially when several images of the same
astronomical scene are combined in order to perform the
measurements. In the context of the Legacy Survey of Space and Time
(LSST) at Vera C. Rubin Observatory \citep{LSST09}, we consider 
two cosmological probes: the measurement of distant 
Type Ia Supernova (SN) 
lightcurves (see \citealt{Astier11} for a review), and the measurement of galaxy shapes (or more precisely
quantities derived from second moments) for evaluating cosmic
shear (see \citealt{Mandelbaum18} for a review). The inferred quantities, respectively flux and
shape, depend on the determination of the position (see \citealt{Guy10} for flux and \citealt{Refregier12} for shape). In both cases, the
noise in the position estimation generally biases the estimator of flux or
second moments. For  cosmic shear tomography (i.e., evaluating
the shear correlations in redshift slices as a function of redshift;
see e.g.  \citealt{Troxel18} or \citealt{Hikage19}) a bias depending on signal-to-noise level translates into a
redshift-dependent bias, potentially disastrous for evaluating
cosmological constraints, in particular regarding the evolution with
redshift of structure formation \citep{Refregier12}. In the same vein, a bias affecting
supernova fluxes in a redshift-dependent fashion compromises the
expansion history derived from the distance-redshift relation \citep{Guy10}.

For repeated imaging of the same sky area, the issue of
position uncertainties inducing measurement biases can be mitigated by
using a source position common to all images: this common
position is less affected by noise than positions measured
independently on individual images, in particular in the context of Rubin Observatory, 
where two back-to-back 15-s exposures are the current baseline observing plan \citep{LSST19}. 
However, averaging positions over images requires accurate
mappings between image coordinate systems, or equivalently mappings from image
coordinate systems to some common frame. In the case of galaxy shape
measurements, one could rely on coadding images prior to the measurement
itself; again, this requires accurate coordinate mappings.

We have mentioned above the bias in the flux or shape estimate caused
by the noise in the position estimate. A bias in the position estimate
also biases the flux or shape measurement.  If biases in position estimates are
spatially correlated, this induces a spatial correlation pattern
between shape estimates. Spatially
correlated biases in shape are clearly a concern because the correlation
function of shear is the prime observable of cosmic shear.

For ground-based wide-field imaging, atmospheric
turbulence contributes to the astrometric uncertainty budget, in
particular for Rubin Observatory observing mode, which consists of two back-to-back 15-s exposures:
distortions induced by the atmosphere appear to scale empirically as
$T_{exp}^{-1/2}$ \citep[\citetalias{Bernstein17} hereafter]{Heymans-CFHT-12, Bernstein17}, where $T_{exp}$ is the
integration time of an exposure. This turbulence contribution
correlates measured positions in an anisotropic fashion (as we will
show later), with a spatial correlation pattern that varies  from
exposure to exposure.

If shape measurements are performed on co-added images, the
astrometric residuals will affect the measured shapes of the galaxies
in a correlated way, and the point spread function (PSF) of the
co-added image in the same way.  The challenge here is to properly
account for the complex correlation pattern for PSF shape parameters,
induced by the combination of anisotropic components of varying
orientation and correlation length. The PSF of the co-added image has two components, 
one due to the actual PSF of the individual exposures, and one due to misregistration, 
in particular, due to turbulence-induced position shifts. Since all input images do 
not contribute equally over the co-added image area, it is common to transform all 
input images to the same PSF prior to co-adding, so that masked areas and gaps 
between sensors do not cause PSF discontinuities on the sum. This PSF homogenization 
does not cope with misregistration, which will then contribute small PSF discontinuities 
on the co-added image.  In the framework of
measurements performed on individual exposures but relying on a common
position estimate, one could perhaps devise a scheme to evaluate the
shape correlations introduced by correlated position residuals, but
this would likely require a significant effort to attain the required
accuracy. For the case of measuring light curves of distant
supernovae, one can readily evaluate the size of systematic position
residuals for a given exposure and correct the flux estimator for the
induced bias.

If one aims to measure galaxy shapes on a large series of short
exposures, reducing the atmospheric contribution to astrometric residuals
will improve the usability of shape measurements, mostly because of the complex
correlation pattern of astrometric shifts induced by the atmosphere. As we will discuss later in this paper, reducing these 
astrometric residuals will be necessary for LSST cosmic shear measurements.
Reducing the astrometric systematics biases will also benefit
other science goals of Rubin Observatory -- for example, trans-Neptunian object searches (see \citealt{Bernardinelli20} as an example), 
or the measurement of proper motions of stars too faint to be measured by {\it Gaia} \citep{LSST19}.

In this paper, we investigate bulk trends of astrometric residuals
presumably dominated by atmospheric turbulence, and aim to model these spatial correlations in order to reduce astrometric residuals. 
For this purpose we are using data from the Subaru telescope equipped with the Hyper Suprime-Cam wide-field camera. 
We first describe the Hyper Suprime-Cam, the data set, and the reduction methods in
Sec.~\ref{sec:hsc_presentation}, and justify the probable atmospheric
origin of the observed position residuals (at least for bright
sources, where shot noise does not dominate). We then describe in Sec.~\ref{sec:model_with_gp} the modeling
of the spatial distribution of residuals as anisotropic Gaussian processes. 
In Sec.~\ref{sec:results} we present our results, in particular
the reduction in variances and covariances that the modeling
provides. In Sec.~\ref{Section_mean_function_results}, we average the a posteriori residuals in instrument 
coordinates in order to detect the small position distortions presumably due to sensors. In Sec.~\ref{ref_disc}, we evaluate
  the expected size of turbulence-induced position offsets for Rubin Observatory,  
  and estimate the spurious shear correlations this causes if not corrected, under some reduction scheme
  We conclude in
Sec.~\ref{conclusions}.

While we were finishing this paper, \citealt{fortino2020reducing}  (\citetalias{fortino2020reducing} hereafter)
produced a paper on the same subject using somewhat different techniques, and
using the Dark Energy Survey (DES) dataset. That work uses results from an
astrometric solver described in \citetalias{Bernstein17}, very
similar to ours. The main differences between the two works are due to the
different sites (Cerro Tololo vs.~Mauna Kea), telescope sizes (4 m vs.~8 m),
telescope mounts (equatorial vs.~alt-az), and instruments (in particular, our
instrument is equipped with an atmospheric dispersion corrector). We will
compare our results to \citetalias{fortino2020reducing} when relevant.

\section{Astrometric solution and residuals from the Hyper Suprime-Cam}
\label{sec:hsc_presentation}
Hyper Suprime-Cam (HSC) is a prime-focus imaging instrument installed
on the 8.2-meter Subaru telescope. Its 104 2k$\times$4k deep-depleted CCD sensors
cover about $1.7$ deg$^2$ on the sky. A detailed description of the
instrument can be found in \cite{HSC18}. The public 
data\footnote{Public data can be found at \href{https://hsc.mtk.nao.ac.jp/ssp/data-release/}{https://hsc.mtk.nao.ac.jp/ssp/data-release/}, and the raw images can be downloaded from \href{https://smoka.nao.ac.jp/HSCsearch.jsp}{https://smoka.nao.ac.jp/HSCsearch.jsp}} used in this study
 correspond to observations acquired for the Deep and UltraDeep layers of the Subaru Strategic
Program\footnote{Details of the SSP survey design and implementation can be found at
  \href{https://hsc.mtk.nao.ac.jp/ssp/}{https://hsc.mtk.nao.ac.jp/ssp/}
  .} (SSP). The data consist of a series of exposures of 180 s to 300 s, in each of the five
optical bands of the camera ($grizy$). In a given
night, exposures in a given band are acquired in sequence, with a fixed
orientation of the camera on the sky. The camera is equipped with an
atmospheric dispersion corrector. The data analyzed in this paper correspond to \numbervisitsSSP exposures, imaged in two 
seasons of searches for transient objects using HSC \citep{Yasuda19}: from November 2016 to May 2017 
in the COSMOS field, and from March 2014 to November 2016 in the SXDS field. 

\subsection{Astrometric solution for HSC data}

We reduced the HSC data using classical procedures: simple
overscan and bias subtraction, implementation of a flat-field correction from
exposures acquired from an in-dome screen, detection of sources using
SExtractor \citep{Sex}, complemented by position estimations from a Gaussian profile fit to the detected sources, and an initial solution for the world coordinate system (WCS) determined by
matching the image catalogs to an external catalog (USNO-B), with
typical residuals of 0.1\arcsec.

\begin{figure*}
	\centering
	\includegraphics[scale=0.47]{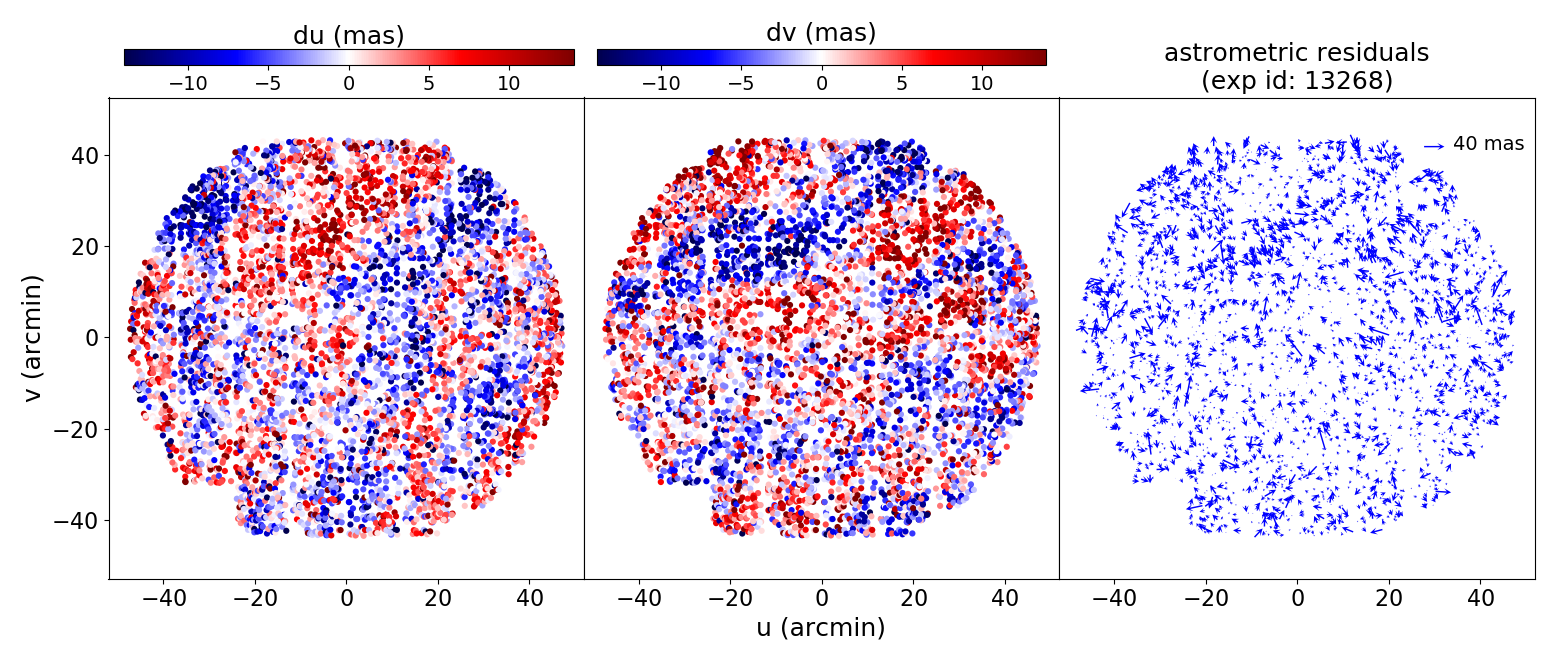}
	\includegraphics[scale=0.47]{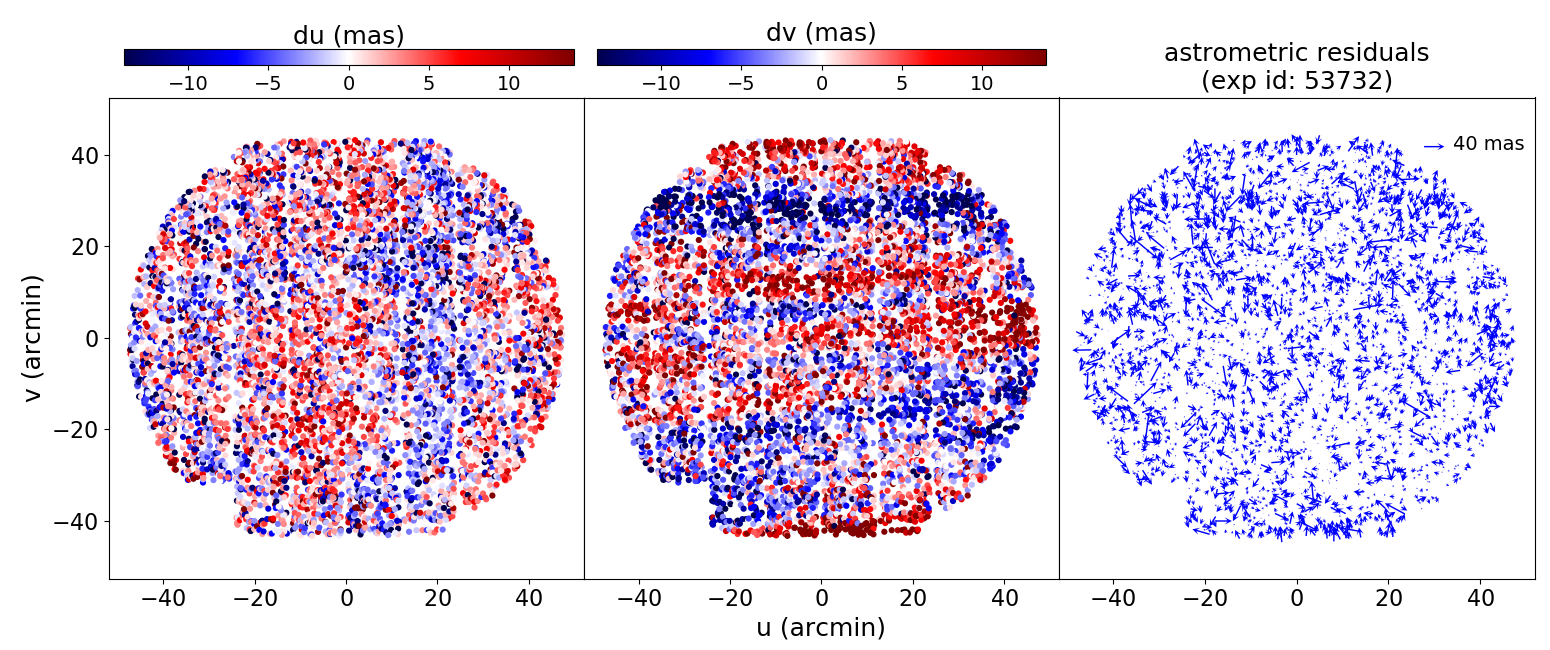}
	\includegraphics[scale=0.47]{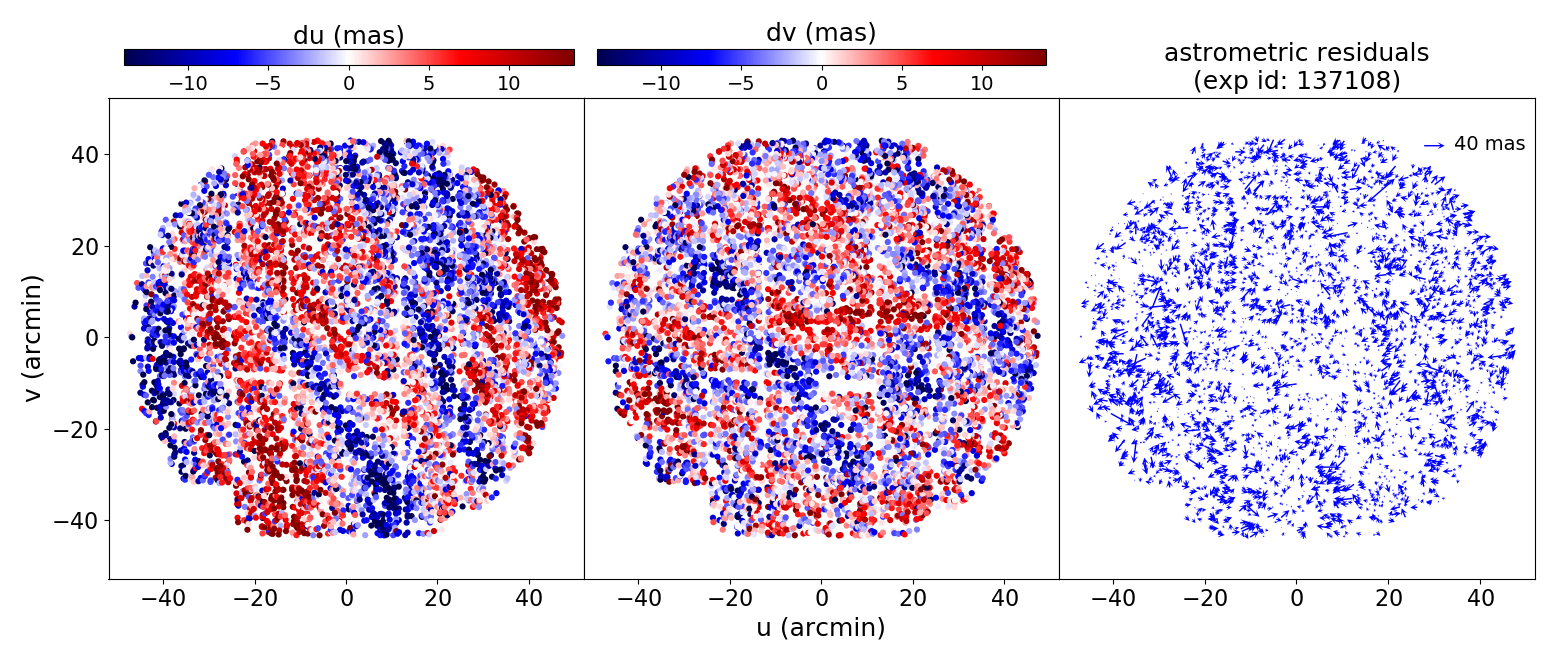}
	\caption{\small Distribution of astrometric residuals in right ascension ($du$, left column) and declination ($dv$, center column), and displayed as vectors (right column) for 
	  three individual exposures in $z$ band, as a function of  position in the focal plane, indexed in 
	arcmin with respect to the optical center of the camera. Gaps in the distributions are due to 
	non-functionning CCD channels.}
	\label{137108_z}
\end{figure*}

We then performed a simultaneous astrometric fit to the catalogs
for the images for each night and each band separately --  5 to 15
images of the same field, with dithers of the order of a few arc
minutes. The WCS's for the input images are used to associate the
detections of the same astronomical sources in different images. We
simultaneously fit the geometrical transformations from pixel space to
sky coordinates, and the coordinates of sources detected in at least  two images. The fit is
 possible because we constrain a small fraction of the detected source positions 
to their Gaia (Data Release 2, \citealt{GaiaDR2}) positions on the date of observation; we call these
sources the ``anchors''. The geometric transformations are modeled as
a per-CCD mapping of pixel coordinates to an intermediate plane
coordinate, followed by a per-image mapping from this intermediate
space to the tangent plane for this specific image. As such, this
model is degenerate and we lift the degeneracy by imposing that one of
the per-image transformations is the identity transformation. The rationale for this
two-transformation model is that the per-CCD transformations capture
the placement of the CCDs in the focal plane and the optical
distortions of the instrument, while the per-image transformations
capture the variations from image to image due to both flexion of the
optics and atmospheric refraction\footnote{SCAMP \citep{SCAMP06} uses
  a  similar parametrization for the same problem. WCSFIT
  \citepalias{Bernstein17} also implements a similar
  parametrization. Our software package, which was developed for Rubin Observatory,
  takes advantage of sparse linear algebra techniques and solves
  for the sources' positions, whereas \citetalias{Bernstein17} chose to treat source positions as
  latent variables, which yields the same estimators.}. We initially model each transformation as a third-order polynomial -- i.e., with 20 coefficients per transformation, or $\sim$2000 parameters
for the instrument geometry, and 20 addtional parameters per image. The typical number of sources in a fit is $\sim$200,000, with a few thousands of these anchored to
Gaia.
The least-squares minimization takes
advantage of the sparse nature of the problem and runs in about 60 s
for the typical number of images (ten). The input uncertainties on the measured positions
 account for shot noise only; therefore, we add a position-error floor of 4 mas to all sources when computing 
the weight used in the least-squares fit to avoid over-fitting
the bright sources. 

The analysis presented in the next section
uses the residuals of this fit as input. The typical rms
deviation of the astrometric residuals for HSC is between 6 and 8
mas for bright sources (with the value depending on the band).  The goal is to reduce the
remaining dispersion and small scale correlations.

\subsection{Astrometric residuals of HSC}

We denote with $du$ and $dv$ the components of the astrometric residual 
field within the local tangent plane, in right ascension and declination, respectively.
In Fig.~\ref{137108_z}, we show examples of the astrometric residuals, 
projected on the tangent plane, for three 
exposures.  At this stage, the stochastic distortions
affecting each exposure are modeled, as described above, with a small number of parameters
(typically 20).  The spatial correlation of the residuals, and the
variability of the correlation from exposure to exposure, indicate
that the  parametrization with a third-order polynomial is not flexible enough to accommodate the
observed variability. We see that the residuals exhibit a
preferential direction that varies from exposure to exposure. This is
very similar to the residuals observed in DES and 
shown in \citetalias{Bernstein17}.

The spatial variations observed in the astrometric residuals 
exhibit preferential directions within an exposure; our hypothesis, as in
 \citetalias{Bernstein17}, is that these anistropies are mostly due to atmospheric
turbulence. In this case (as discussed in \citetalias{Bernstein17}),  the astrometric residual field follows the
gradient of the optical refractive index of the atmosphere in the
telescope beam, averaged over the integration time of the exposure.
The two-point correlation function $\xi$ for the astrometric residual field has most generally two independent components, but only one
if it is a gradient field and is curl free (Helmholtz's theorem). The decomposition into a curl-free $E$-mode
and a divergence-free $B$-mode of vector fields on the plane \citepalias{Bernstein17} allows us to test the above hypothesis.

Following \citetalias{Bernstein17} (see appendix A in particular), we evaluate the two-point correlation functions $\xi_+$, $\xi_-$, and $\xi_{\times}$ for the astrometric residual field,
\begin{align}
\xi_+(r, \beta) &=\left< dX\left(\textbf{X}\right) e^{-i \beta} \left[dX\left(\textbf{X} + \textbf{r} \right) e^{-i \beta}\right]^{\star} \right>, \\
\xi_-(r, \beta)+ i\xi_{\times}(r, \beta)  &=\left< dX\left(\textbf{X}\right) e^{-i \beta}  dX\left(\textbf{X} + \textbf{r} \right) e^{-i \beta} \right>,
\end{align}
where $dX(\textbf{X})$ is the astrometric residual field on a complex form 
($dX = du + i dv$) measured at a position $\textbf{X}$ on the focal plane, $r$ is the distance
between two positions, and $\beta$ is the position angle of $\textbf{r}$.
From $\xi_+$ and $\xi_-$ we find the correlation functions $\xi_E$ and $\xi_B$ correponding to the $E$- and $B$-modes of the
field:
\begin{align}
\xi_{E} &= \frac{1}{2} \xi_+(r) +  \frac{1}{2}\xi_-(r) -  \int_r^{\infty}\frac{1}{r'}\xi_-(r') dr' \\
\xi_{B} &= \frac{1}{2} \xi_+(r) -  \frac{1}{2} \xi_-(r) +  \int_r^{\infty}\frac{1}{r'}\xi_-(r') dr' 
\end{align}

We evaluate the correlation functions by calculating the covariance between the astrometric residuals for pairs of sources, in bins of spatial separation between the sources. From these, 
we compute the binned version of $E$- and $B$-mode correlation functions. 
This involves an integral over separation (Eq.~37 in \citetalias{Bernstein17}) that we evaluate by simply
summing over each bin. 
With this binned estimation method, the integral
of the correlation functions over angular separations is zero (see Appendix \ref{integral_correlation_function}).

\begin{figure}
	\centering
	\includegraphics[scale=0.35]{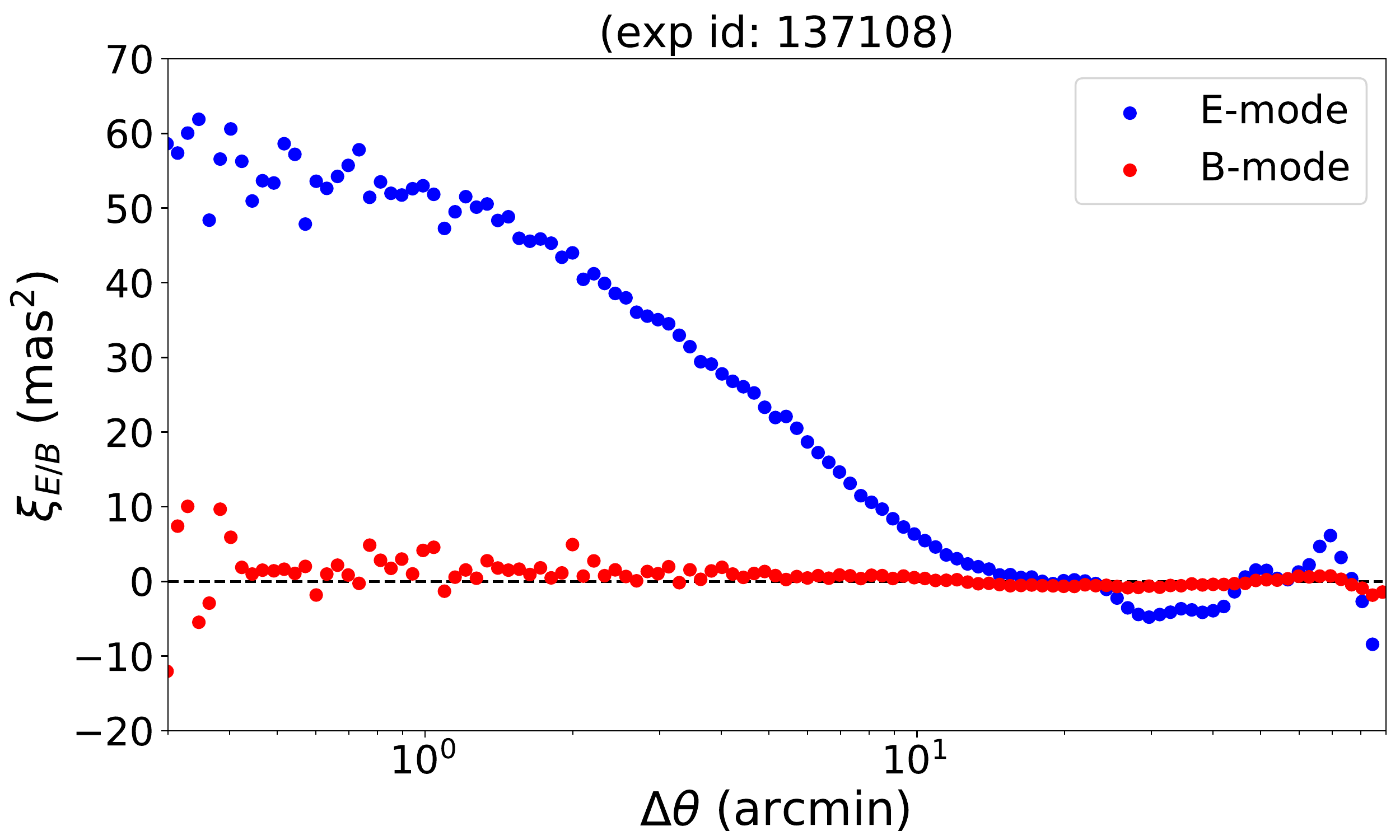}
	\caption{\small Typical $E$- and $B$-mode correlation functions for a single 300-s $z$-band exposure of HSC.}
	\label{137108_z_eb_mode}
\end{figure}

\begin{figure}
	\centering
	\includegraphics[scale=0.29]{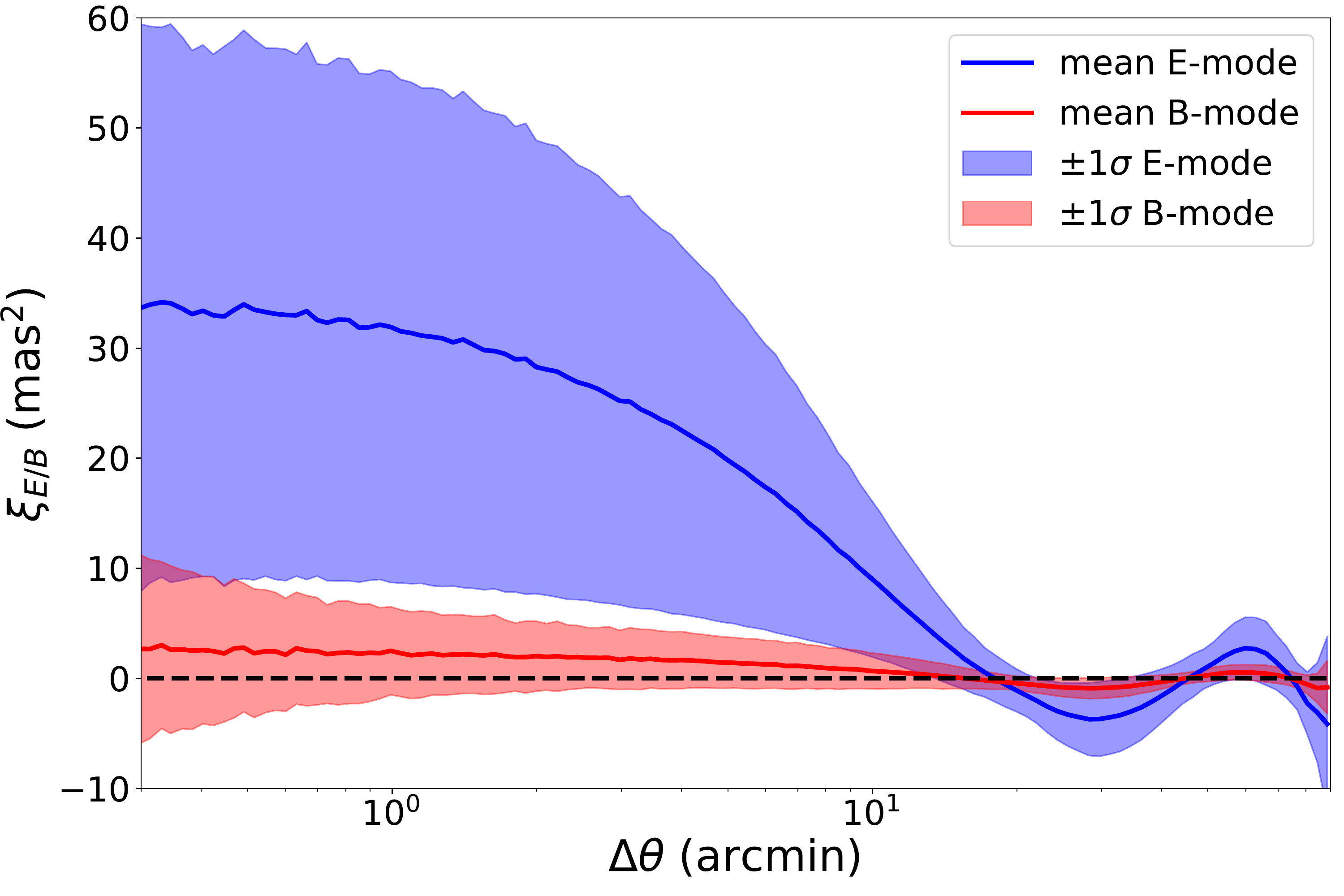}
	\caption{\small Average over all  \numbervisitsSSP exposures of the $E$- and $B$-mode correlation functions. 
	The blue and red shaded area represent respectively the standard deviation across nights of  $E$- and  $B$- mode. }
	\label{eb_mean_all_ssp_vk}
\end{figure}

\begin{figure*}
	\centering
	\includegraphics[scale=0.47]{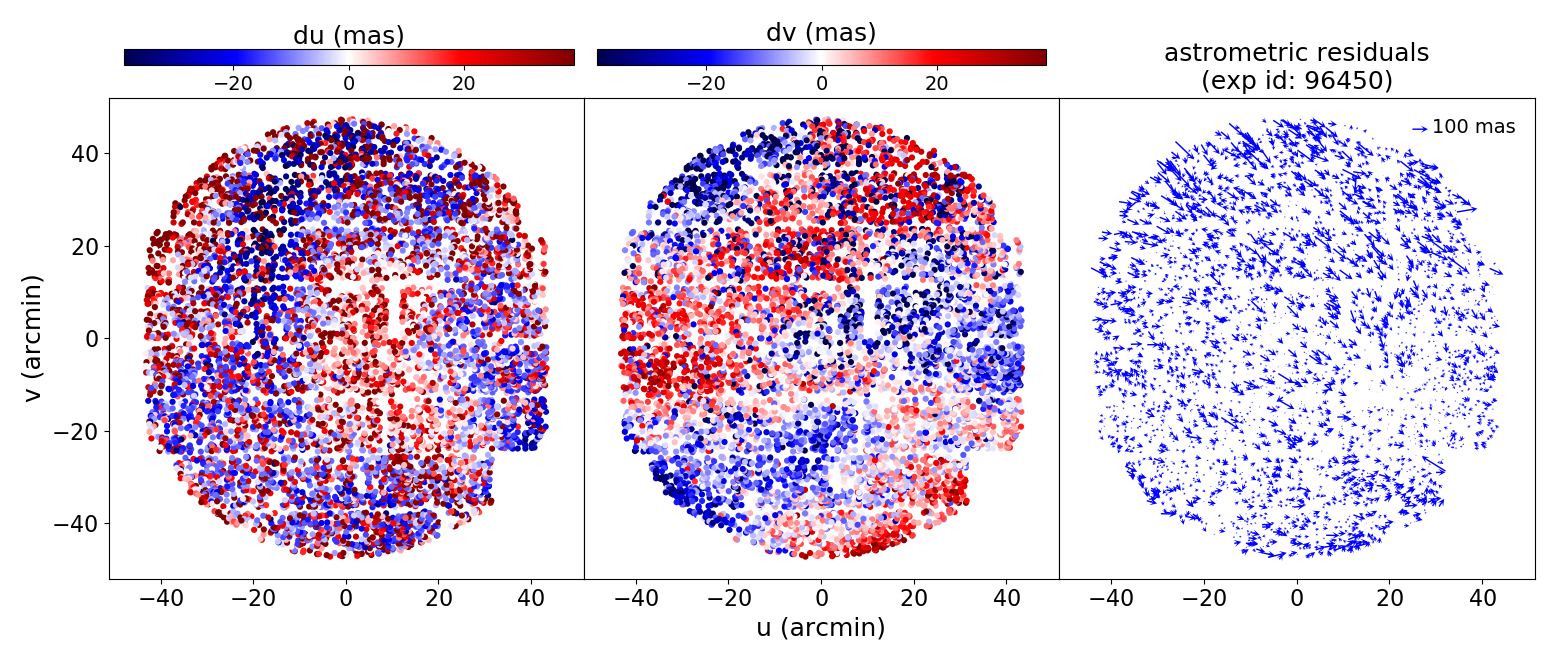}
	\caption{\small Distribution of astrometric residuals in right ascension ($du$, left) and declination ($dv$, center), and displayed as vectors (right), as a function of  position in the focal plane.
	This $z$-band exposure corresponds to one of two nights when the focal plane was rotated through large angles during the course of the night.}
	\label{96450_z_b_mode}
\end{figure*}

In Fig.~\ref{137108_z_eb_mode}, we show the correlation functions corresponding to the 
$E$- and $B$-modes for the single exposure shown in the bottom row of 
Fig.~\ref{137108_z}). Figure~\ref{eb_mean_all_ssp_vk} shows the
mean value calculated over the \numbervisitsSSP exposures of the Deep and  UltraDeep layers of the SSP. We observe that the $E$-mode correlation function is non-zero while the $B$-mode correlation function is
compatible with zero. This is consistent with the result reported by DES in \citetalias{Bernstein17}, for
a different observing site and with a significantly different
instrument. The HSC and DES results both indicate that the deplacement field can be described as the gradient
of a scalar field, which is likely the average over the line-of-sight
of the optical refractive index of the atmosphere, as argued in \citetalias{Bernstein17}. 
We also note that the measured correlation function has negative values at large
separations; this is unavoidable, because the integral of the correlation
function is zero (see Appendix \ref{integral_correlation_function}). 
Moreover, the correlations that can be described by 
 the fit of a third-order polynomial over the exposure data are
much smaller than the observed correlation, and should have a correlation length of
about 0.4$^\circ$\footnote{Since the size of the field of view is 1.7~deg$^2$, and 10 parameters per component are fit for the third-order polynomial,  each parameter describes an area of $\sim$0.17~deg$^2$, which corresponds to an angular scale of $\sim$0.4$^\circ$}.

\subsection{Difference between HSC and DES results}
A notable difference between our results and those reported in \citetalias{Bernstein17} is the value of the correlation function as the angular separation approaches 0: Fig.~11 in \citetalias{Bernstein17},
indicates a low-separation covariance of $E$-modes of the order of $80
\ \text{milli-arcsec}^2$ while we observe a value of $\rmsastroHSC
\ \text{milli-arcsec}^2$ for HSC; see Fig.~\ref{eb_mean_all_ssp_vk}.
The fitting algorithms used to calculate the residuals are similar. However, 
there are two important quantitative differences: the exposure times are longer for HSC ($\sim$270~s vs.~90~s for DES), and the mirror diameter of the
Subaru telescope is about twice as large as that of the Blanco Telescope (where DES observed). The variance attributed to atmospheric effects typically scales
as the inverse of the exposure time \citep{Heymans-CFHT-12}, while
the dependence on aperture is more complicated, but favors larger
apertures. Finally, the Subaru telescope is located at an elevation of
4400 m at the summit of Mauna Kea in Hawaii, while the Blanco telescope is at an elevation of only 2200 m at the Cerro Tololo Inter-American Observatory (CTIO) in Chile.

\subsection{Nights with non zeros $B$-mode}
\begin{figure}
	\centering
	\includegraphics[scale=0.35]{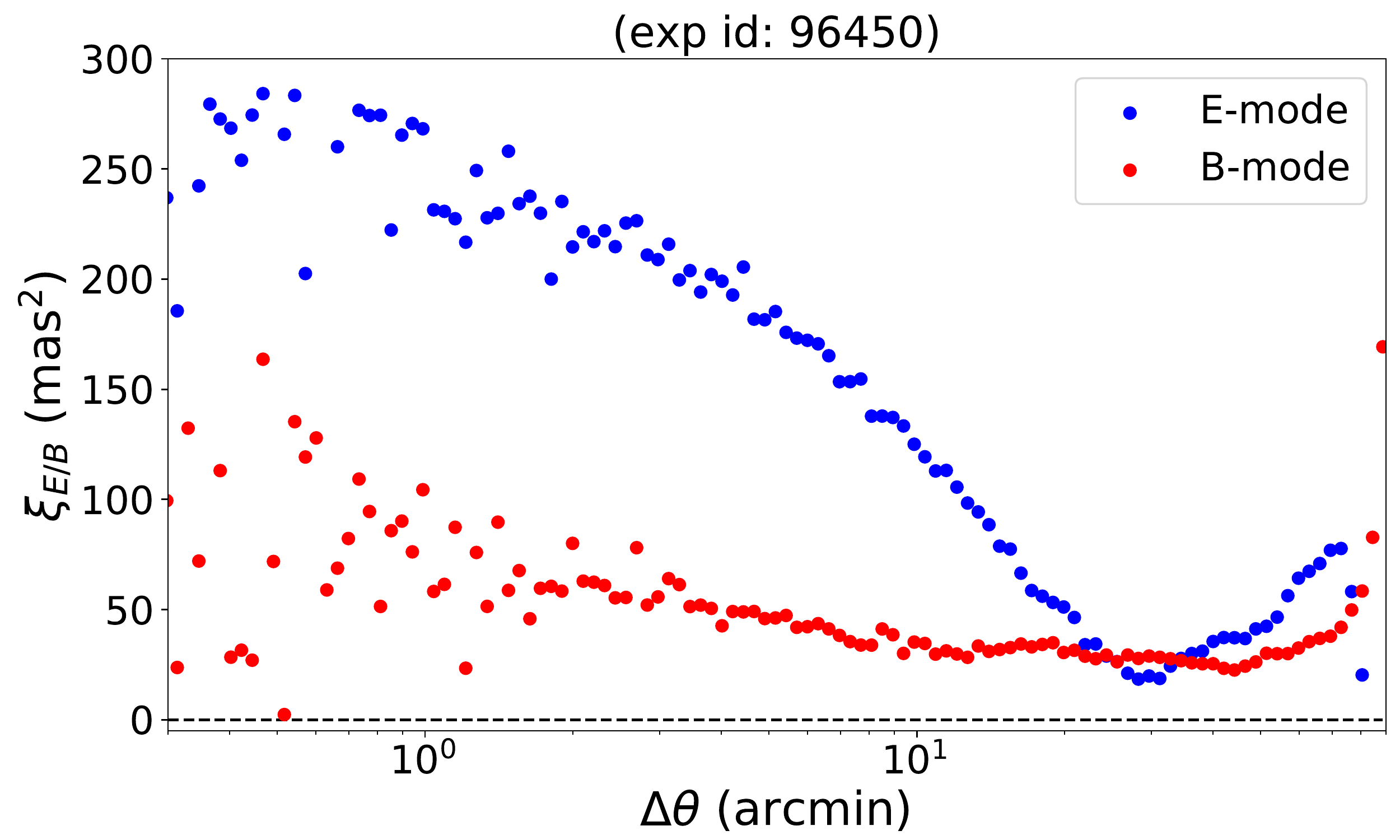}
	\caption{\small Typical $E$- and $B$-mode correlation function for the same exposure shown in Fig.~\ref{96450_z_b_mode}, corresponding to one of two nights when the focal plane was rotated through large angles.}
	\label{96450_z_b_mode_eb_mode}
\end{figure}

Another significant difference between the HSC results reported here and those of DES are that the astrometric residuals in exposures for two HSC nights exhibit correlation functions with
$B$-mode contributions that are not negligible compared to the $E$-mode contributions, and $E$-mode values much larger than typical.
An example of the residuals for an exposure in such a
night is shown in Fig.~\ref{96450_z_b_mode} and the corresponding correlation
functions in Fig.~\ref{96450_z_b_mode_eb_mode}. The Subaru telescope
resides on an alt-azimuth mount; the field derotator mechanism 
rotates only the focal plane, while the wide-field corrector remains fixed with
respect to the telescope.
  Our astrometric model indexes the optical distortions of the
  imaging system with respect to coordinates in CCD pixel space.
  Since in HSC, the image corrector rotates with respect to the CCD
  mosaic, any breaking of the rotational symmetry of the optical
  distortions will cause spurious astrometric residuals. As these
  residual only appear if exposures with different rotation angles are
  fit together, they tend to grow with the range of rotation angles
  involved in the fit. These residuals are not induced by the
  gradient of a scalar field, and hence are prone to similar amounts of
  $E$- and $B$-modes.
The nights with large $B$-mode contributions are characterized
by a large rotation of the focal plane over the course of the observations, mostly because the observations were spread over several hours, and sometimes most of the night.
  Our astrometric model does not compensate for this rotation (and neither does
the one in \citetalias{Bernstein17}), both because we were not aware of the details
of the HSC mechanics before discovering these large rotations, and
 because the astrometry software we are using was originally developed
for reducing the Canada-France-Hawaii Telescope (CFHT) Legacy Survey,
and the CFHT has an equatorial mount. The fitted model would have to be heavily modified to account for these rotations, as noted in \citetalias{Bernstein17}, for only the two night impacted by these large rotations.
Most of our fits include images acquired over less than an hour and the whole rotation range is typically less than 20$^\circ$.

Other than these two nights with large focal plane
rotations, the astrometric residuals exhibit a covariance pattern that
can be mostly attributed to atmospheric turbulence. In the next section, we describe how we
model these residuals.

\section{Modeling astrometric residuals using anisotropic Gaussian processes}
\label{sec:model_with_gp}
Spatial variations in the refractive index of the atmosphere can be described as a
Gaussian random field, which must be stationary (that is, the covariance
between values at different points depends only on their separation)
because there is no special point in the image plane. We therefore model the  astrometric residual field in the image
plane as a Gaussian process (GP), which allows us to correct for 
 astrometric residuals in the data, accounting for the correlations introduced by the Gaussian random field.
In Sec.~\ref{section_gp_intro}, we briefly introduce 
GPs, describe how the astrometric residual field can be modeled as a GP, and  
discuss possible strategies for optimization of the GP. In  
Sec.~\ref{section_hyp_opt}, we describe the method we use for the optimization of the 
GP hyperparameters, and our choice for the analytical correlation function. This GP  
interpolation was originaly developped for interpolating the atmospheric part of the 
PSF in the context of DES \citep{Jarvis21} and follow a similar 
scheme of interpolation.

\subsection{Introduction to Gaussian processes}
\label{section_gp_intro}

In this subsection, we give a brief overview of classical
GP interpolation\footnote{Also referred to as ordinary
  kriging or as Wiener filtering.}; a more detailed description can
be found, for example, in \cite{Rasmussen06}. A GP is the
optimal method for interpolating a Gaussian random field. GP interpolation can be used for irregularly spaced datasets and, because it operates by describing the correlations between data rather than following a specific functional form\footnote{This feature of GPs results in the descriptor "non-parametric," which simply means GPs are not from a parametric family of distributions. As with any interpolation method, a GP requires setting parameter values.}, is very flexible. In practice, one must choose both the analytic form of this correlation function (also called a "kernel") and the parameters for this kernel (often called "hyperparameters") that describe the second-order statistics of the data.

A stationary Gaussian random field is entirely defined by its mean value
(as a function of position) and its second-order statistics, which depend only on separation in position. 
We denote the (scalar) Gaussian random field as $y$ 
and its mean value as $y_0 \equiv E[y]$, where $E[x]$ denotes the expectation value of the random variable $x$. Because the field is Gaussian (i.e., $y$ is Gaussian distributed at any position $X$), the distribution is Gaussian:
\begin{equation}
  y \sim {\cal N}(y_0, \xi(0)),
  \label{real_gp_equation}
\end{equation}
where $\xi$ is the correlation function: 
\begin{equation}
\xi(X'-X) \equiv {\rm Cov}(y(X), y(X')).
\end{equation}

In the context of modeling astrometric residuals, $y$ can be either of the components 
$du$ or $dv$, and $X$ is the coordinate in the local tangent plane.
GP interpolation is a method for estimating the value of the field at
arbitrary locations, given a realization of the field at a set of (usually
different) locations. In practice, the kernel 
$\xi$ is unknown -- or at least its parameters are unknown.  We will
discuss their determination in the next section. 
The covariance matrix for the data realization is calculated and used in the
interpolation. The covariance matrix will be positive definite (which is required for it to be invertible in a later step) for any set of locations if and only if the correlation function $\xi$ has a positive Fourier transform (Bochner's Theorem). This
constrains the shape of possible correlation functions.

We now describe the practical interpolation method. We have
a realization $y_i$ at positions $X_i$, and we assume for now that we know 
the kernel $\xi$. The covariance matrix $\bs{C}$ of the $\bs{y}$ realization is given by:
\begin{equation}
  C_{ij} \equiv {\rm Cov}[y_i-y_0(X_i)), y_j-y_0(X_j)] = \xi(X_j-X_i) + \delta_{ij} \sigma_i^2 , \label{eq:cov_training_sample}
\end{equation}
where $\sigma_i$ is the measurement uncertainty of $y_i$.

The expectation of the Gaussian field at locations $\bs{y'}$, given the values
of $\bs{y}$ at locations $\bs{X'}$, is \citep{Rasmussen06}:
\begin{equation}
\label{gp_intep_equation}
\bs{y'} = \boldsymbol{\Xi}(\bs{X'}, \bs{y}) \bs{C}^{-1} (\bs{y}-\bs{y_0}) + \bs{y_0},
\end{equation}
where $\boldsymbol{\Xi}$ is a matrix with elements defined by
\begin{equation}
  \Xi(\bs{A}, \bs{B})_{ij} \equiv \xi(\bs{A}_i - \bs{B}_j) .
  \label{gp_estimator}
\end{equation}
The covariance of the interpolated values is:
\begin{multline}
  E \left[ (\bs{y'}-\bs{y_0}(\bs{X'})) (\bs{y'}-\bs{y_0}(\bs{X'}))^T \right ] = \\ \boldsymbol{\Xi}(\bs{X'}, \bs{X'}) -\boldsymbol{\Xi}(\bs{X'}, \bs{X})^T  \bs{C}^{-1} \boldsymbol{\Xi}(\bs{X'}, \bs{X}) .
\end{multline}

We see from Eq.~\ref{gp_intep_equation} that, in the absence of
measurement uncertainties $\sigma_i$, the interpolated values at the
training points $\bs{X}$ are just the $\bs{y}$ values with no
uncertainties. This is because the interpolation method delivers the
average expected field values given $\bs{y}(\bs{X})$ with
covariance $\bs{C}$. In practice, the matrix
$\bs{C}$ (defined in Eq.~\ref {eq:cov_training_sample}) is 
numerically singular or almost singular if there are no measurement
uncertainties $\sigma_i$, even for sample sizes as small as 20; in
the case of zero uncertainty, a small noise value should be added for the above
expressions to be numerically stable.

The interpolation method is now well-defined, given a data realization and a choice of kernel.
A commonly used kernel
is the Gaussian kernel (also known as a squared exponential),
\begin{equation}
\label{RBF_kernel}
\xi \left(\textbf{X}_1, \textbf{X}_2\right) = \phi^2 \ \exp\left[-\frac{1}{2}
\left(\textbf{X}_1 - \textbf{X}_2 \right)^T \textbf{L}^{-1} \left(\textbf{X}_1 - \textbf{X}_2 \right) \right] ,
\end{equation}
where $\bs{X}_1$ and $\bs{X}_2$ correspond 
to two positions (in the focal plane for our case),  
$\phi^2$ is the variance of the Gaussian random field 
about the mean function $\bs{y_0}(\textbf{X})$, and  
the covariance matrix $\textbf{L}$ is in general anisotropic since   
atmospheric turbulence typically has a preferred direction due to wind direction:
\begin{equation}
\textbf{L} = \left(\begin{matrix}
                \cos\alpha & \sin\alpha\\
                -\sin\alpha & \cos\alpha\\
            \end{matrix}\right)^T
            \left(\begin{matrix}
                \ell^2 & 0\\
                0 & (q \ \ell)^2\\
            \end{matrix}\right)
            \left(\begin{matrix}
                \cos\alpha & \sin\alpha\\
                -\sin\alpha & \cos\alpha\\
            \end{matrix}\right) ,
\end{equation}
where $\alpha$ represents the direction of the anisotropy, $\ell$ represents the correlation length in the isotropic case, $q$ is the ratio of the semi-major to semi-minor axes 
of the ellipse associated with the covariance matrix\footnote{Can be written as 
a function of the image distortion parameters $g_1$ and $g_2$. This parametrization 
of the covariance matrix is analogous to that used in cosmic shear; see, for example,
equation 7 in \citealt{Schneider05}.}. 

Although a Gaussian kernel is often used, we can 
choose an analytical form for the kernel based on empirical considerations and/or the relevant 
physics. Both the functional form of the kernel and the hyperparameters associated with the kernel determine the
interpolated estimates. Once the kernel shape is chosen, the parameters can be determined 
with a maximum likelihood fit, where the
likelihood of the realization $\bs{y}$ (of size $N$) is defined by
\begin{equation}
 {\cal L}= \frac{1}{{\sqrt{2\pi}}^N} \times  
 \frac{1}{\sqrt{\det(\bs{C})}} \times %\\ 
 \exp\left(-\frac{1}{2} \left(\textbf{y} - \textbf{y}_0 \right)^T \bs{C}^{-1}\left(\textbf{y} - \textbf{y}_0 \right)\right) .
 \label{max_likelihood_GP_hyper}
 \end{equation}
 
Maximizing this expression with respect to the parameters of $\xi$ that define
$\bs{C}$ (via Eq.~\ref{eq:cov_training_sample}) is numerically
cumbersome because it involves many inversions (in practice, factorizations) of the covariance matrix $\bs{C}$, which has the size of the
``training sample'' $\textbf{y}$.  The factorisation (for example,
the Cholesky decomposition) also trivially delivers the needed determinant. The time to
compute such a factorisation scales as
${\mathcal{O}}\left(N^3\right)$, where $N$ is the size of the training
sample $\bs{y}$.  It is  possible to speed up the inversion
of the matrix $\bs{C}$ in Eq.~\ref{max_likelihood_GP_hyper} under
certain assumptions. For example, \citealt{GEORGE} propose to
speed-up the matrix inversion from ${\mathcal{O}}\left(N^3\right)$ to
${\mathcal{O}}\left(N \log^2(N)\right)$ based on a special
decomposition of the matrix (HODLR).  Other methods achieve 
${\mathcal{O}}\left(N\right)$ under certain assumptions about the
analytical form of the kernel and by being limited to
one dimension \citep{CELERITE}.

Here, we follow another route, which relies on the good sampling
provided by the training sample. From the smoothness of the measured correlation function in  Fig.~\ref{137108_z_eb_mode}, 
we can conclude that the average distance to the nearest neighbors is much
smaller than the correlation length.
Therefore, we can estimate the anisotropic correlation function directly from the data. The practical implementation is described in the next section.

\begin{figure}
	\centering
	\includegraphics[scale=0.59]{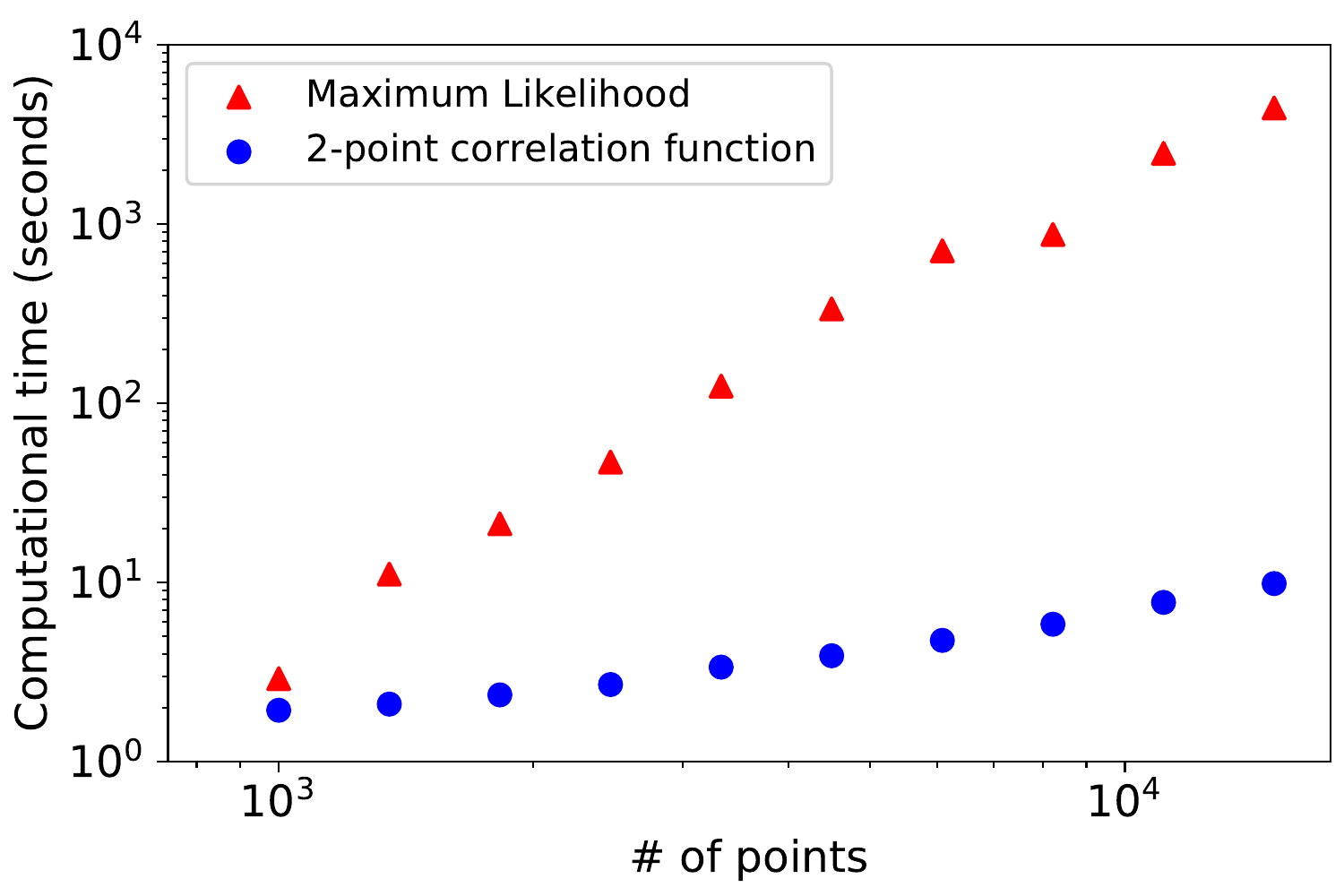}
	\caption{\small Typical computational time necessary to determine hyperparameters as a function of the number of data points used during training for a classical maximum likelihood method using Cholesky decomposition (red triangles) and when using the two-point correlation function computed by TreeCorr (blue circles). The typical number of training points for astrometric residuals modeled using GPs is between $\sim10^3$ and $\sim10^4$.}
	\label{time_hyp_search}
\end{figure}

\subsection{Hyperparameter estimation using the two-point correlation function}
\label{section_hyp_opt}

Estimating the kernel of a stationary GP directly from
the two-point statistics was pioneered in the field of geostatistics 
(see \citealt{Cressie92} for example).
The  time to compute the two-point correlation function naively scales as ${\mathcal{O}}\left(N^2\right)$; however,
faster approaches have been developed. We use a package called TreeCorr
\citep{Jarvis04}, which evaluates covariances in
distance bins for large datasets. Our implementation of the estimation of hyperparameters 
using TreeCorr can be found online\footnote{\GitLinkTreegp}. 
The computational time for TreeCorr depends on the bin size (see section 4.1 of \citealt{Jarvis04}), and it proves
particularly efficient for  data sets of a size relevant to PSF interpolation, where it scales roughly linearly with
the number of input data points. Figure~\ref{time_hyp_search} shows the
computational time for the maximum likelihood approach and the 
estimation of hyperparameters based on the binned
two-point correlation function using TreeCorr, as a function of the number of data points; the latter technique is several
 orders of magnitude faster for the training sample sizes we are contemplating (between $\sim 10^3$ and $\sim 10^4$).

We estimate the covariance matrix of the binned two-point 
correlation function via a bootstrap (done on sources). 
This measured covariance matrix is then used in the fit of the analytical model for the 
kernel to the measured two-point correlation function in a 
non-diagonal least-squares minimization.
These three steps (binned two-point correlation function, 
bootstrap covariance matrix,
least-squares fit) are included in the computational times 
shown in Fig.~\ref{time_hyp_search}.

\begin{figure}
	\centering
	\includegraphics[scale=0.7]{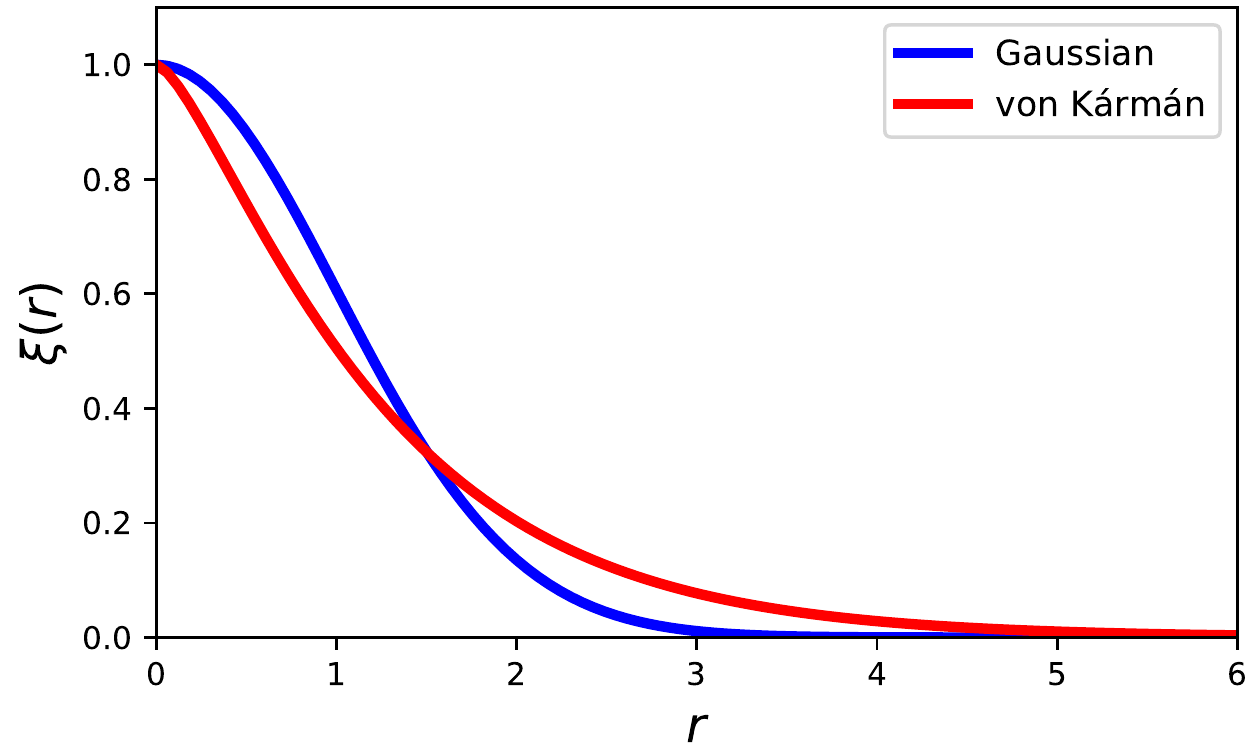}
	\caption{\small Gaussian kernel compared to the von K\'arm\'an kernel
          that is used in this analysis. The width of the latter was determined by a least-squares fit to the former. }
	\label{rbf_vs_vk_isotropic}
\end{figure}

Although the Gaussian kernel is often used for GP interpolation, 
for ground-based imaging, a kernel profile with broader wings is expected to provide a better description of the longer-range correlations present in PSF dominated by atmospheric turbulence
(see, for example, Fig.~2 in \citealt{RODDIER81}).
To account for the clear anisotropy in the correlation function, we use an anisotropic von K\'arm\'an kernel as proposed in \citealt{Heymans-CFHT-12} to describe the observed spatial correlations of
PSF distortions for CFHT and is parametrised as
\begin{multline}
\label{VK_kernel}
\xi \left(\textbf{X}_i, \textbf{X}_j\right) = \phi^2  \times \left[ \left(\textbf{X}_i - \textbf{X}_j \right)^T \textbf{L}^{-1} \left(\textbf{X}_i - \textbf{X}_j \right) \right]^{\frac{5}{6}} \times 
 \\ K_{-5/6} \left[ 2 \pi \left(\textbf{X}_i - \textbf{X}_j \right)^T \textbf{L}^{-1} \left(\textbf{X}_i - \textbf{X}_j \right) \right],
\end{multline}
where the notation is the same as for Eq.~\ref{RBF_kernel} and $K$
is the modified Bessel function of the second kind.  At large
separations, $\xi$ decays exponentially. We show in
Fig.~\ref{rbf_vs_vk_isotropic} Gaussian and von K\'arm\'an kernels of
similar widths. As we will soon show, a von K\'arm\'an kernel
reduces the residuals more efficiently than a Gaussian kernel. 

\begin{figure*}
	\centering
	\includegraphics[scale=0.53]{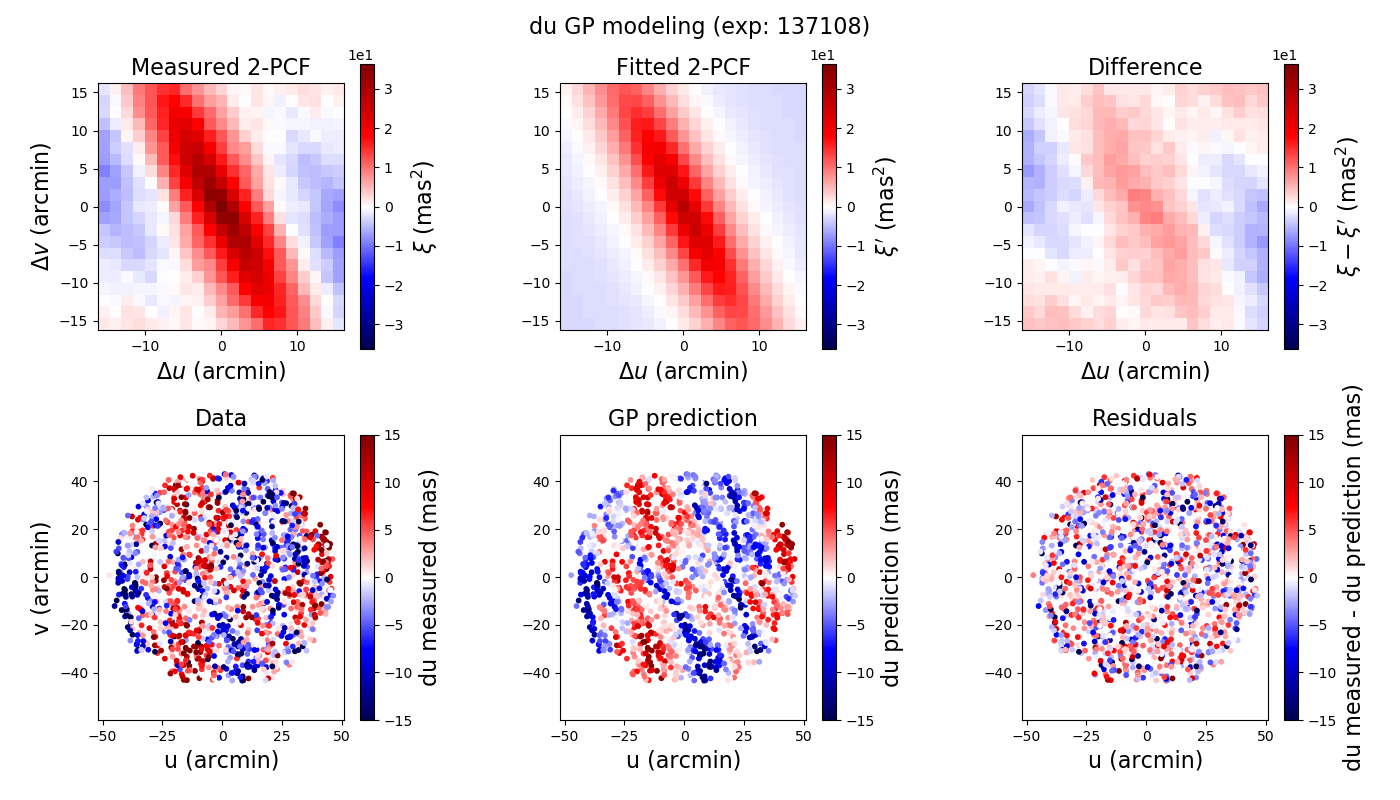}\\
	\includegraphics[scale=0.53]{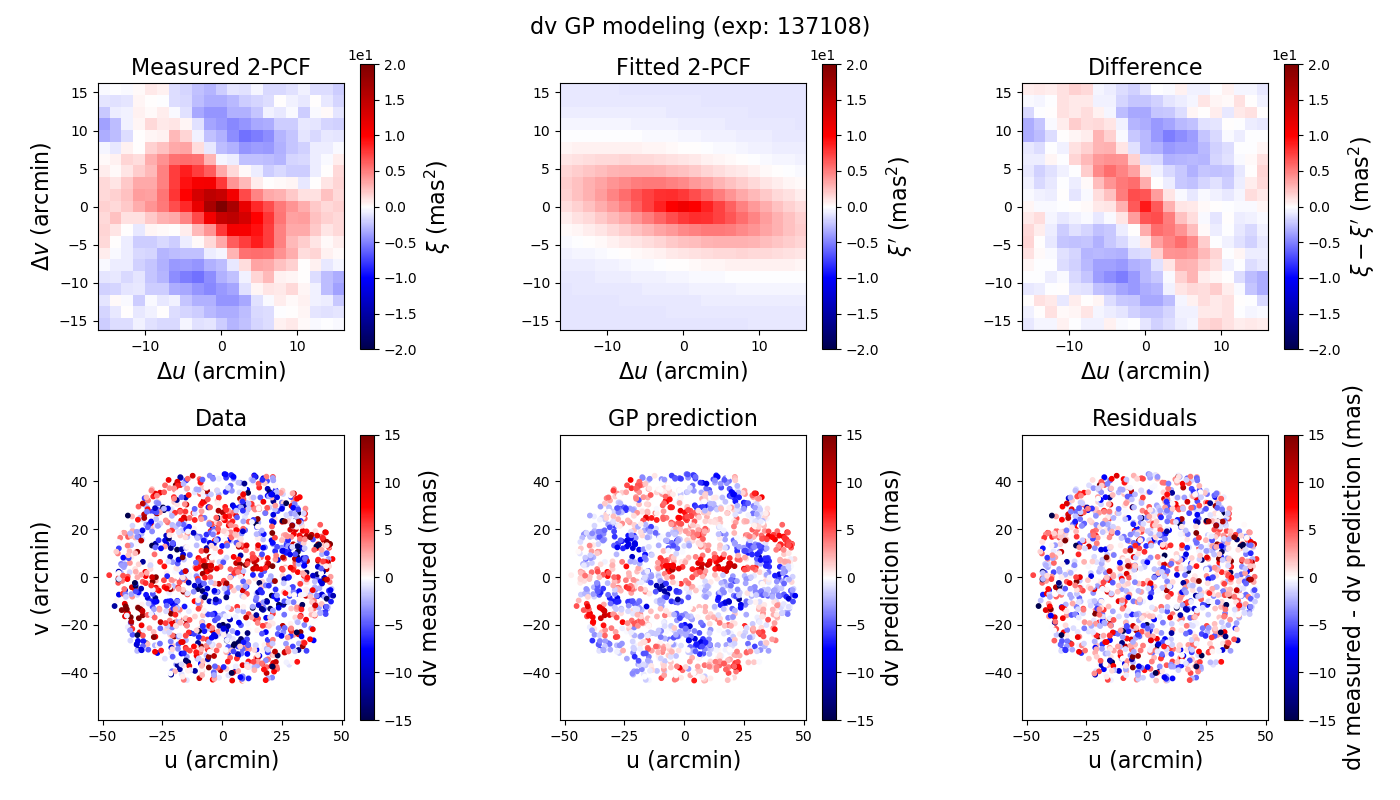}\\
	\caption{\small GP fit of astrometric residuals for a single 300-s $z$-band exposure. The fit is done independently for each component of the vector field. The top six plots show results for the $du$ component, and the bottom six plots for the $dv$ component. The plots in the top row within each group illustrate how the hyperparameters are determined from a von K\'arm\'an kernel (center plot) fit to the measured two-point correlation function calculated for 80\% of the training sample (left plot); the difference between the measured two-point correlation function and the best-fit von K\'arm\'an kernel is shown in the right-most plots. The plots in the bottom row within each group of six represent the variation across the focal plane of the measured component of the astrometric residual field, the astrometric residual field predicted by the GP using the best-fit hyperparameters, and the difference between measure and fit, projected in the local tangent plane for the 20\% of sources in the validation sample.}
	\label{results_one_exp}
\end{figure*}

One may wonder why we rely on a parametrized form of the kernel,
typically with a small number of parameters (four here), rather than
using a more empirical fit (for example with spline functions) of the
measured correlation function. As discussed earlier, a correlation
function should have a positive Fourier transform and, if it does not, the
covariance matrix of observations (Eq.~\ref{eq:cov_training_sample}
for $\sigma_i=0$) is not positive-definite for all realizations. Therefore,
when smoothing the measured correlation function, one should restrict
the outcome to functions with positive Fourier transforms. 
Splines cannot in general be guaranteed to have positive Fourier transforms. Both Gaussian and von K\'arm\'an kernels fulfill this requirement, so we only consider these two models. We note that a maximum likelihood approach
faces the same constraint of having a correlation function with a positive Fourier transforms.

\begin{figure}
	\centering
	\includegraphics[scale=0.35]{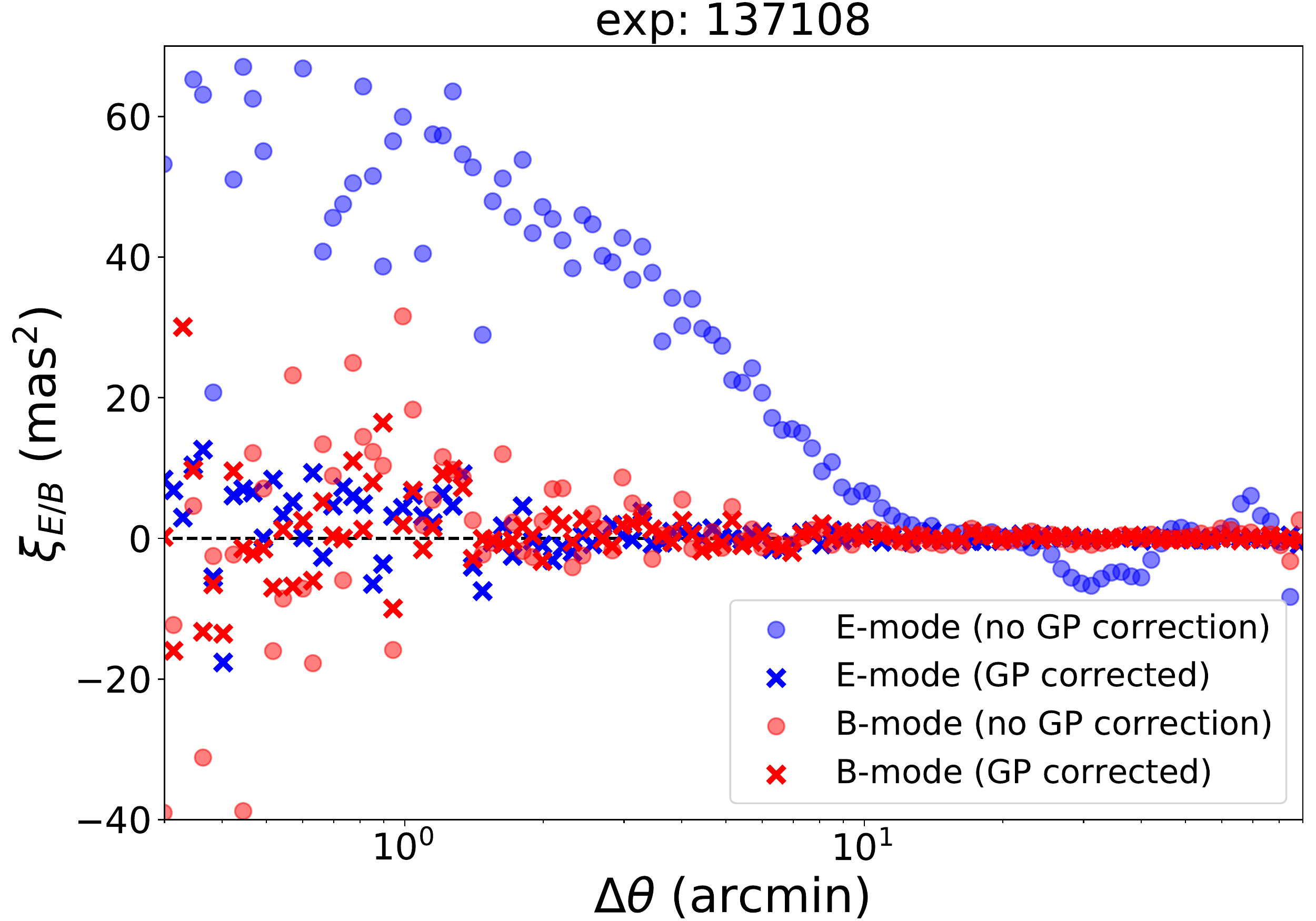}
	\caption{\small Correlation functions corresponding to  $E$- (blue)
          and $B$- (red) modes of the astrometric residual field on the validation sample for a representative exposure (one of three exposure shown in Fig.~\ref{137108_z}),
          before applying the GP interpolation correction (circles) and after correction (crosses)}
	\label{results_one_exp_eb_mode}
\end{figure}

\begin{figure}
	\centering
	\includegraphics[scale=0.3]{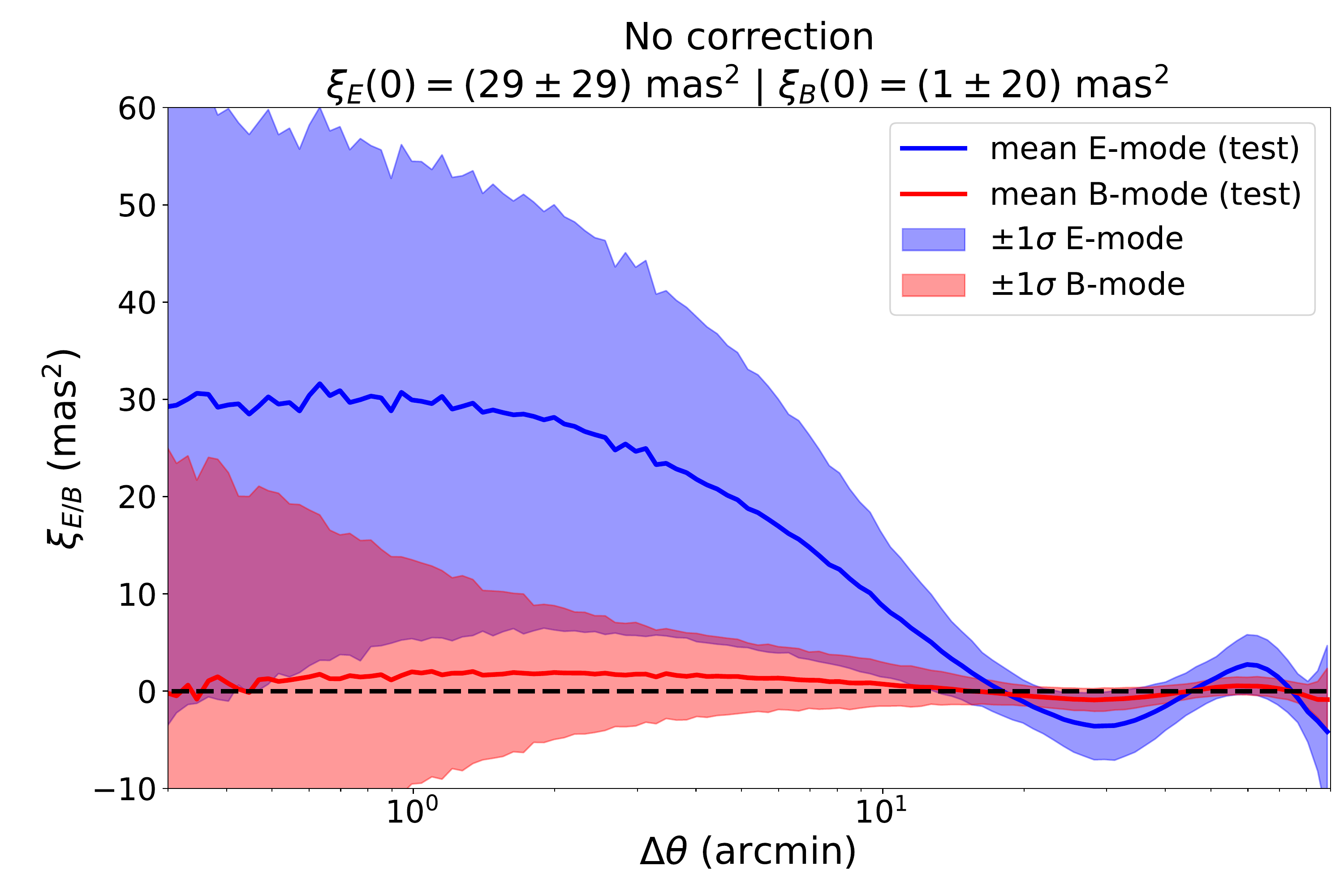}\\
	\includegraphics[scale=0.3]{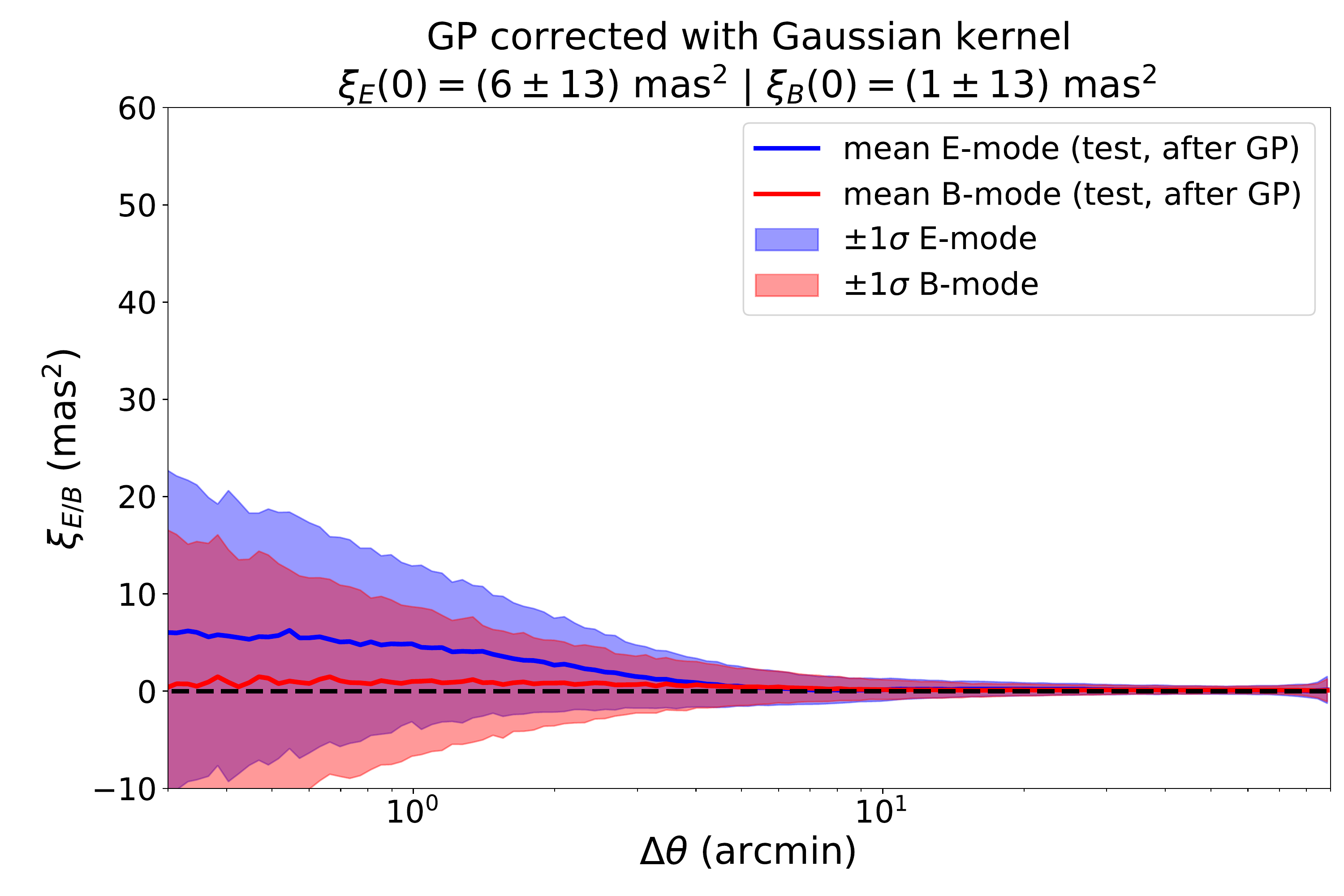}\\
	\includegraphics[scale=0.3]{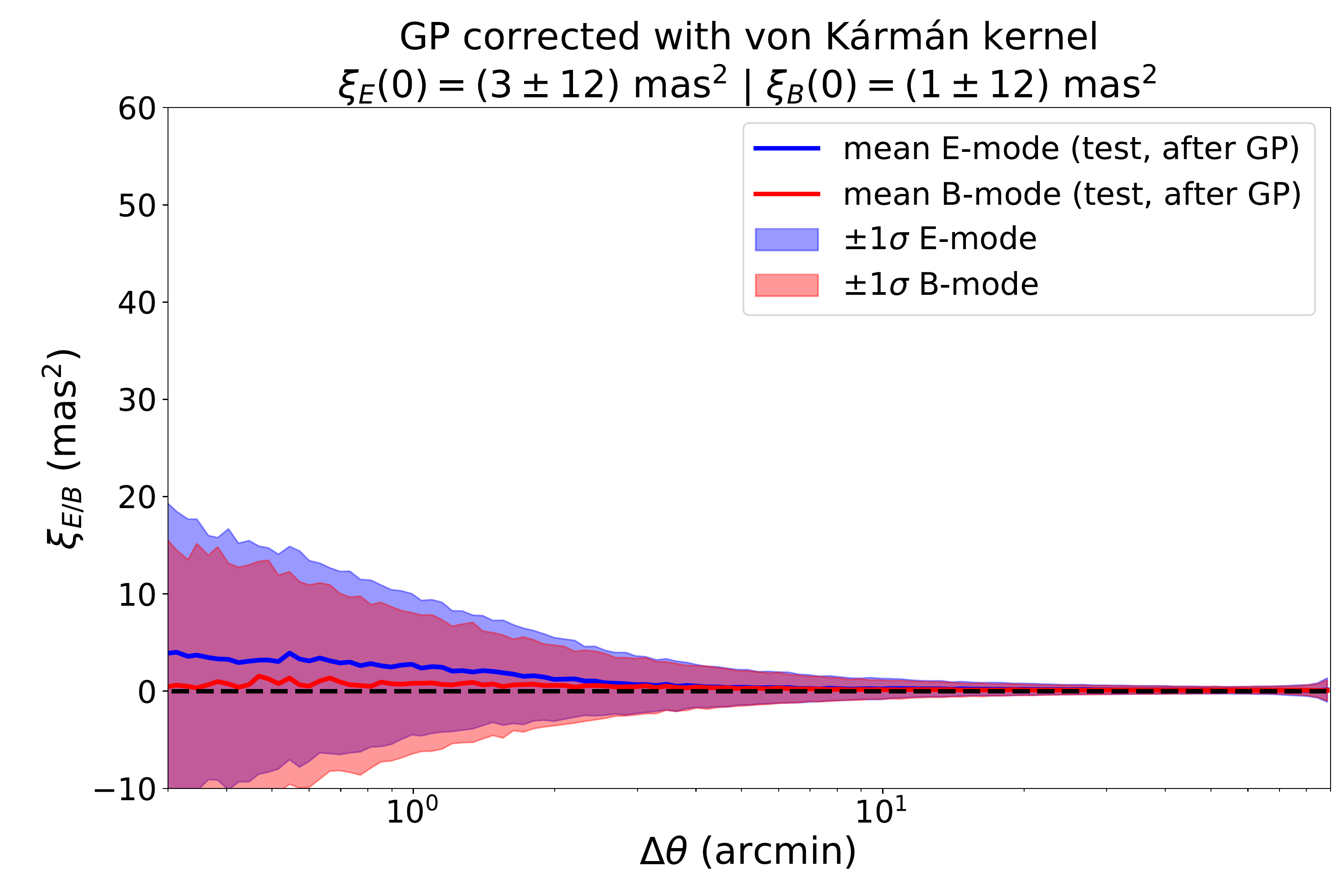}
	\caption{\small Average over all  \numbervisitsSSP exposures of the $E$- and $B$-mode correlation functions, calculated on the validation sample, for different modeling choices. 
	Top: Correction from GP interpolation is not applied. Middle: correction from GP interpolation is applied using a Gaussian kernel. 
	Bottom: correction from GP interpolation is applied using a von K\'arm\'an kernel.  For all those plots, the blue and red shaded 
	area represent respectively the standard deviation across nights of  $E$- and  $B$- mode. }
	\label{residuals_gp_eb}
\end{figure}

\subsection{Difference with DES GP modeling}
\label{section_F20}

We now describe the main differences with the approach chosen in \citetalias{fortino2020reducing}.
First, we do not enforce the model to describe a gradient field; rather we
 model and fit independently the two spatial components. In 
 \citetalias{fortino2020reducing}, the authors eventually optimize the kernel
hyperparameters in order to minimize the average covariance on small
angular scales, while we simply fit the hyperparameters to the
empirical covariance. We speculate whether this optimization is necessary and whether it 
will be viable when it comes to large scale production as
required by the processing of Rubin Observatory data. 
Second, the anisotropy of the correlations in 
\citetalias{fortino2020reducing} is entirely attributed to the
wind-driven motion of a static phase screen during the
exposure, and we have not tested whether this assumption improves our
results. Third, We model the correlation function with four parameters, while
\citetalias{fortino2020reducing} use five, where the fifth parameter is 
the outer scale and we expect its influence on the correlation function to be small.
Finaly, we have not
rejected outliers in the GP fit; i.e., sources
exhibiting large a posteriori residuals are not removed from the
training sample, but a priori outliers have been removed. 

\section{Correcting astrometric residuals using anisotropic Gaussian processes}
\label{sec:results}

We apply GP interpolation to the astrometric
residuals for the \numbervisitsSSP exposures of the SSP in the five bands. We train
the GP on the unsaturated sources with magnitudes brighter than \magmagic AB mag. 
The residuals are clipped at 5-$\sigma$ during the prior astrometric fitting process but, 
in contrast with \citetalias{fortino2020reducing}, we do not clip in the GP fit. We fit the
two projections of the residuals as independent GPs, and ignore cross correlations.

As discussed earlier, GPs simply reproduce the input data if
interpolation at an input data point is requested, at least in the absence
of measurement noise. Therefore, in order to use residuals to evaluate 
the GP performance in an
unbiased way, we randomly select a ``validation sample'' consisting of
20\% of the sources fully qualified for training, that we exclude
from the training sample. We then compute the GP interpolation for
all sources used in the astrometry (i.e., all sources with an
aperture flux delivering a signal-to-noise ratio greater than 10). The
performance tests described in this section (unless otherwise specified) are
 computed only on this validation sample.

The result of the GP modeling of the astrometric
residual field for a single exposure in $z$-band is
shown in Fig.~\ref{results_one_exp} (see caption for detail). The measured correlation functions have negative lobes because they integrate to zero. The positive analytical function cannot obviously reproduce
those negative parts. We add a constant floor $k$ in order to allow the kernel 
to have negative lobes and to not bias hyperparameters estimation. Consequently, the following 
equation is minimized in order to find the best set of hyperparameters $\boldsymbol{\theta}$:

\begin{equation}
\chi^2 = \left(\boldsymbol{\xi} - \hat{\boldsymbol{\xi}}(\boldsymbol{\theta}) -k \right)^T \bs{W} \left(\boldsymbol{\xi} - \hat{\boldsymbol{\xi}}(\boldsymbol{\theta}) -k \right) \ ,
\end{equation}

\noindent where $\boldsymbol{\xi}$ is the measured two-point correlation function, $\hat{\boldsymbol{\xi}}$ is the analytical kernel (gaussian or von K\'arm\'an), 
$\bs{W}$ is the inverse of the covariance matrix of the measured two-point correlation function, and $k$ which is the 
constant not taken into account when computing the final kernel 
and consequently when computing the GP interpolation (Eq.~\ref{gp_intep_equation}).
The von K\'arm\'an kernel fits well the principal direction of the anisotropy and delivers a 
reasonable correlation length. However, one can see that by looking at the two-point 
correlation function residuals on the third row of Fig. ~\ref{results_one_exp} that the
 von K\'arm\'an profile does not fit perfectly the observed kernel, even if it does a better 
 job than the classic Gaussian kernel (see below). A room for improvement would be 
 to have a more flexible kernel to describe the observed two-point correlation, 
 such as a spline basis\footnote{We tried to fit the observed two-point 
 correlation function using a spline basis, but it did not produce a positive definite covariance matrix of the residuals. 
 A solution might be to fit a spline basis on the Fourier transform of the observed 
 two-point correlation function instead and ensure that the fit is positive in Fourier space. 
 This procedure should produce a positive definite covariance matrix of residuals (Bochner's Theorem).}, 
 but lies beyond the scope of this analysis.

We show in Fig.~\ref{results_one_exp_eb_mode} the
correlation functions for the $E$- and $B$-modes of the
astrometric residual field calculated for the validation sample in a representative exposure (same as the one presented in Fig.~\ref{results_one_exp}), before and after GP modeling and interpolation with a von K\'arm\'an kernel.
We see a significant reduction in $E$-mode correlation function after GP interpolation. 

\begin{figure*}
	\centering
	\includegraphics[scale=0.38]{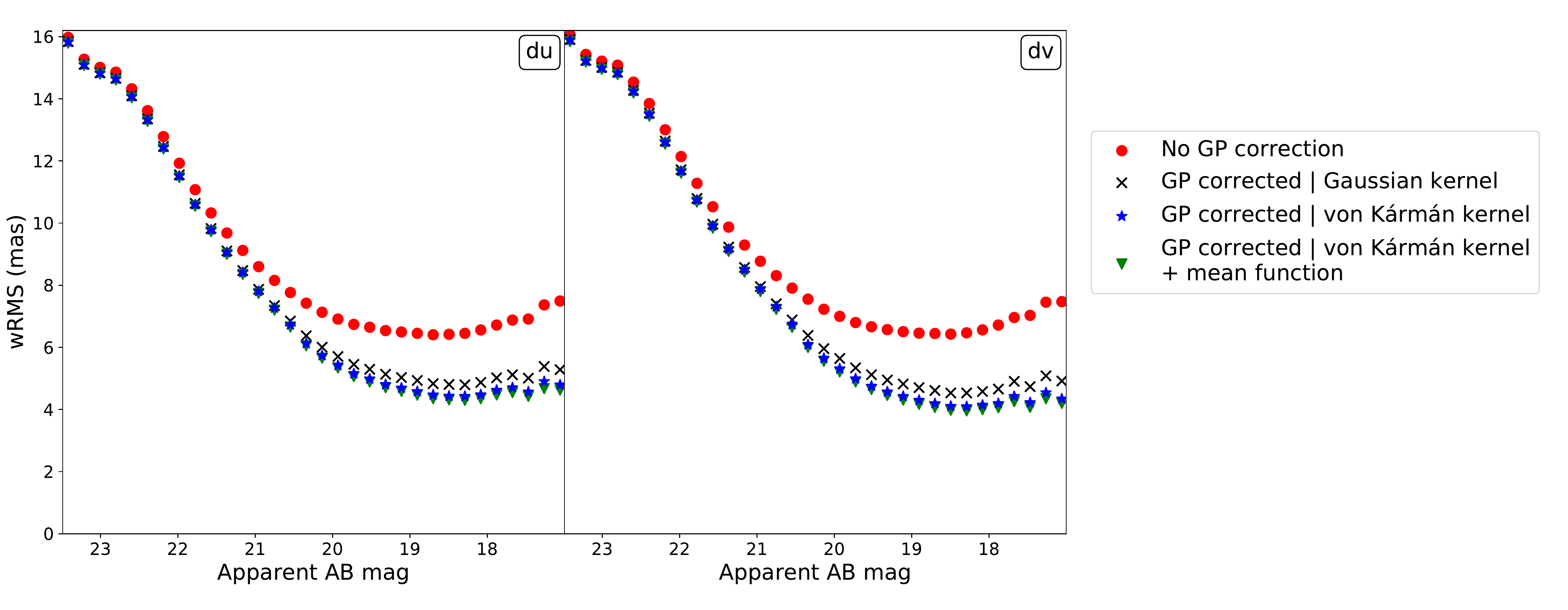}
	\caption{\small Weighted RMS of astrometric residuals as a function of apparent AB magnitude of the source 
	for each component of the astrometric residual field. Four models are compared: no 
	correction from a GP interpolation (red circles), correction from a GP interpolation with a Gaussian 
	kernel (black crosses), correction from a GP interpolation with a von K\'arm\'an 
	kernel (blue stars), and correction from a GP interpolation  with a von K\'arm\'an 
	kernel and the mean function computed as described in Sec.~\ref{Section_mean_function_results} (green triangles, hidden underneath the blue ones).}
	\label{wrms_vs_mag}
\end{figure*}

We show in Fig.~\ref{residuals_gp_eb} the  $E$ and $B$-mode
correlation functions calculated for the validation sample and averaged over all  \numbervisitsSSP exposures, before (top plot) and after GP interpolation with a Gaussian kernel (center plot) and a von K\'arm\'an kernel (bottom plot), together with
the $\pm 1$-$\sigma$ spread over exposures. GP interpolation
reduces the magnitude of the correlation function by almost one order of magnitude, from 30 mas$^2$ to 3 mas$^2$ at small scales.
The von K\'arm\'an kernel performs better than the Gaussian kernel. 
We can also see the improvement by comparing the dispersion of the astrometric residuals 
with the flux as in Fig.~\ref{wrms_vs_mag}. As for the $E$-mode reduction, it can be seen that after correcting by the GP interpolation, GP 
reduces the dispersion by a factor of two and the von K\'arm\'an kernel does a better job than the classical 
Gaussian kernel by $\sim 20 \%$ in rms.

We investigated whether the a posteriori residuals are sensitive
to the size of the training set, and found no compelling differences
between results for exposures with 2000 and 6000 training sources. We interpret this as
an indication that the modeling is not limited by the image plane sampling
or the shot noise affecting position measurements, 
but rather by the ability of the kernel to describe the correlation.
Therefore, reducing the number of training points is an immediate
avenue to reducing computational demands, and further work should focus on  
modeling of the kernel.

\section{Average residuals in CCD coordinates}
\label{Section_mean_function_results}

\begin{figure}
	\centering
	\includegraphics[scale=0.4]{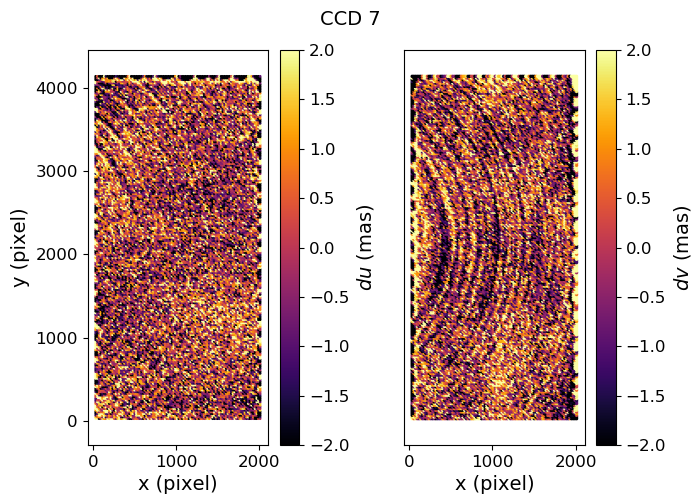}\\
	\includegraphics[scale=0.4]{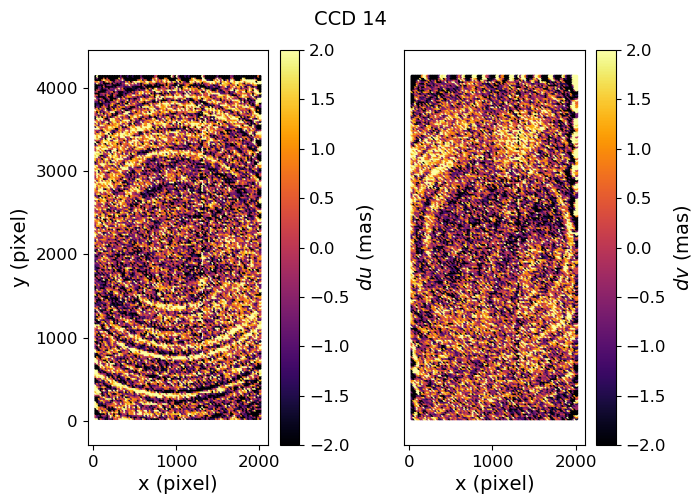}\\
	\includegraphics[scale=0.4]{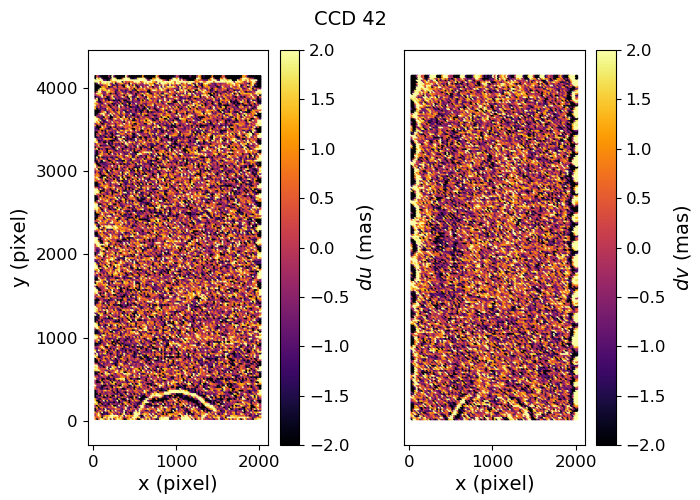}
	\caption{\small Average of the astrometric residual field for each component projected in pixel coordinate 
	for 3 different chips with some characteristic features that can be also found on the other chips.}
	\label{mean_function_some_ccds}
\end{figure}

Sensor effects that are not modeled with the GP 
(e.g., fabrication defects in the CCD, or distortions in the drift fields) 
can appear in an image of the average residuals over the focal plane.
In Fig.~\ref{mean_function_some_ccds}, we show the
average value over all the exposures of the two components of the residuals for three representative CCDs. One can
see two main types of defects: so-called "tree rings", and
"scallop-shaped"  structures near the edges of the sensors. The former are
commonly attributed to the variation in the density of impurities,
with a symmetry due to how the
crystals are grown. The latter are likely due to mechanical stresses applied
to the silicon lattice, induced by the binding of the sensor to its
support structure. The DECam sensors exhibit similar
defects, as shown in \citetalias{Bernstein17}, but at about an order of
magnitude larger scale. The sizes of the tree-ring residuals are $\sim 2$ mas for HSC  and $\sim 13$-26 mas for DECam \citep{Plazas14}. 
We have not been able to find in the literature previous mention of these tiny defects on HSC sensors. 
The rms astronomical residuals associated with the scallop-shaped features in the images for the HSC CCDs are $\sim$10 mas -- larger than those associated with the tree-ring features. 

\begin{figure*}
	\centering
	\includegraphics[scale=0.4]{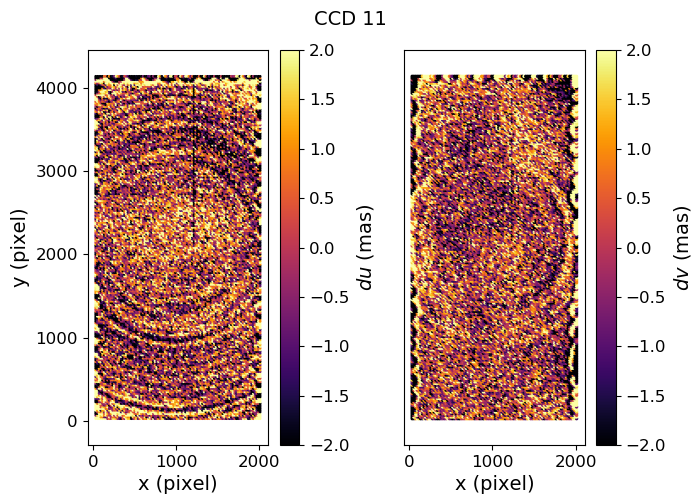}
	\includegraphics[scale=0.4]{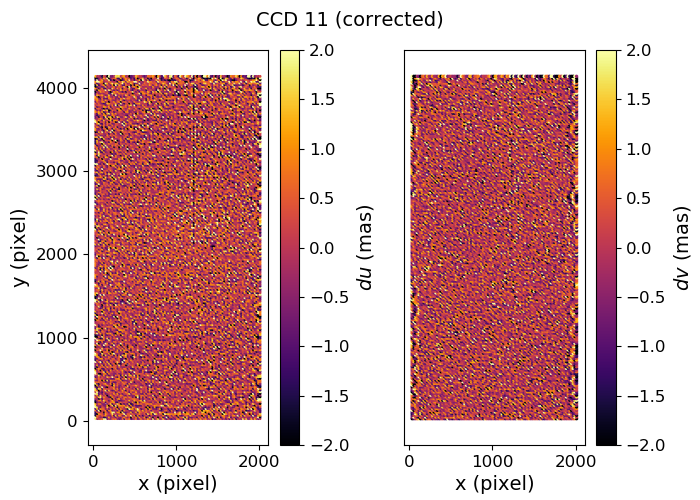}
	\caption{\small The average of the astrometric residual field for each component projected in pixel coordinates for a given chip, 
	before and after including the average in the GP model as a mean function.}
	\label{mean_after_correction}
\end{figure*}

In order to clearly see these small effects (mostly below 10 mas), we
averaged all bands together since the signal in individual bands is
 very noisy. We expect that the patterns in each
band will  differ only on a global scale due to the variation with wavelength of the average conversion depth of photons in silicon. We
mildly confirm that redder bands have weaker patterns, as expected,
but we could not devise a robust measurement of ratios between bands.
Averaging in larger spatial bins smears the smaller structures.
If we apply these patterns as a mean function of the GP,
the improvement of the residuals (both in variance and covariance) is
below 1 mas$^2$ (cf. Fig.~\ref{wrms_vs_mag}). However, 
recomputing the mean function and taking it into account in the GP modeling 
allows us to remove most of the patterns observed in individuals CCDs as shown in Fig.~\ref{mean_after_correction}.

\begin{figure*}
	\centering
	\includegraphics[scale=0.25]{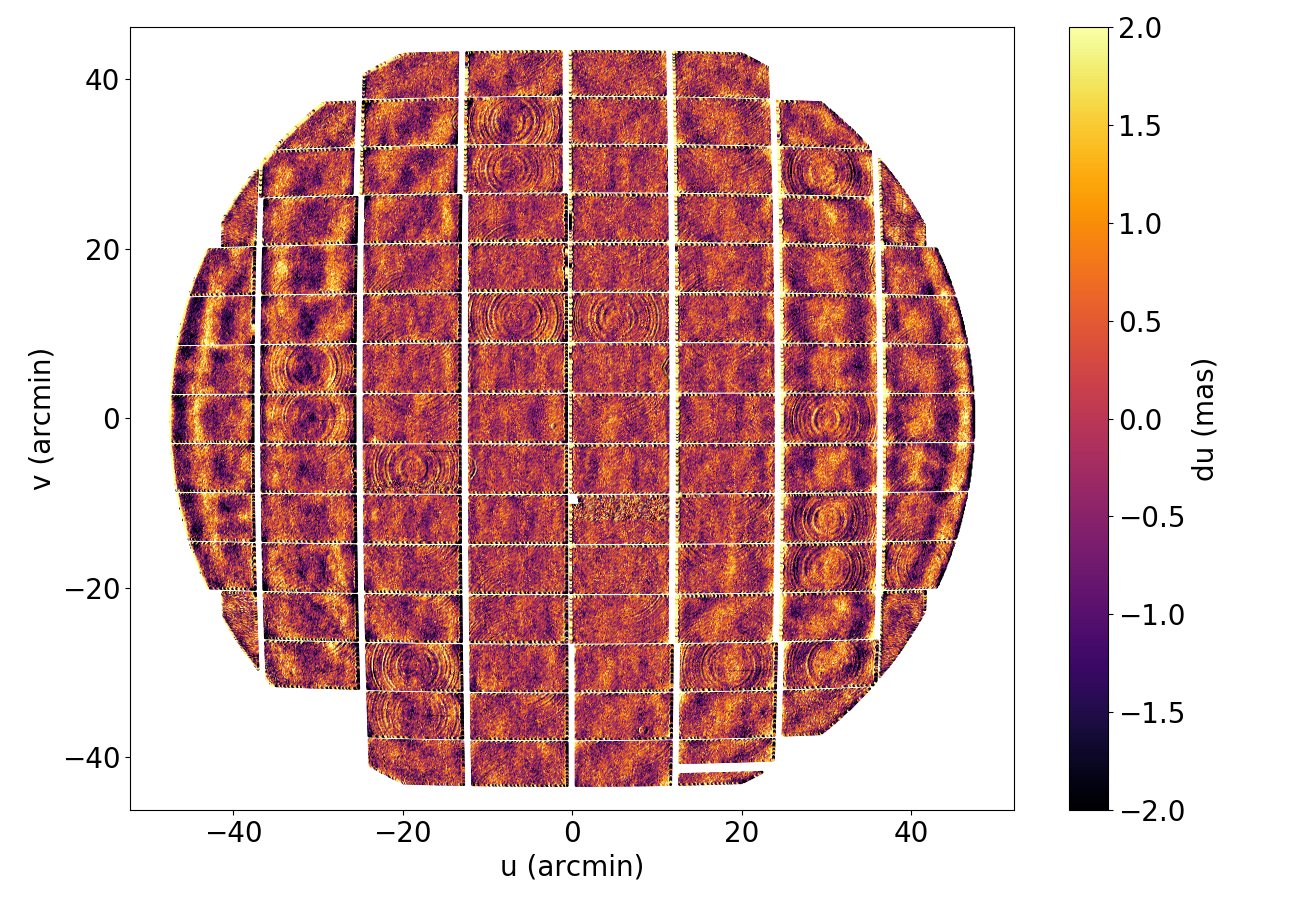}
	\includegraphics[scale=0.25]{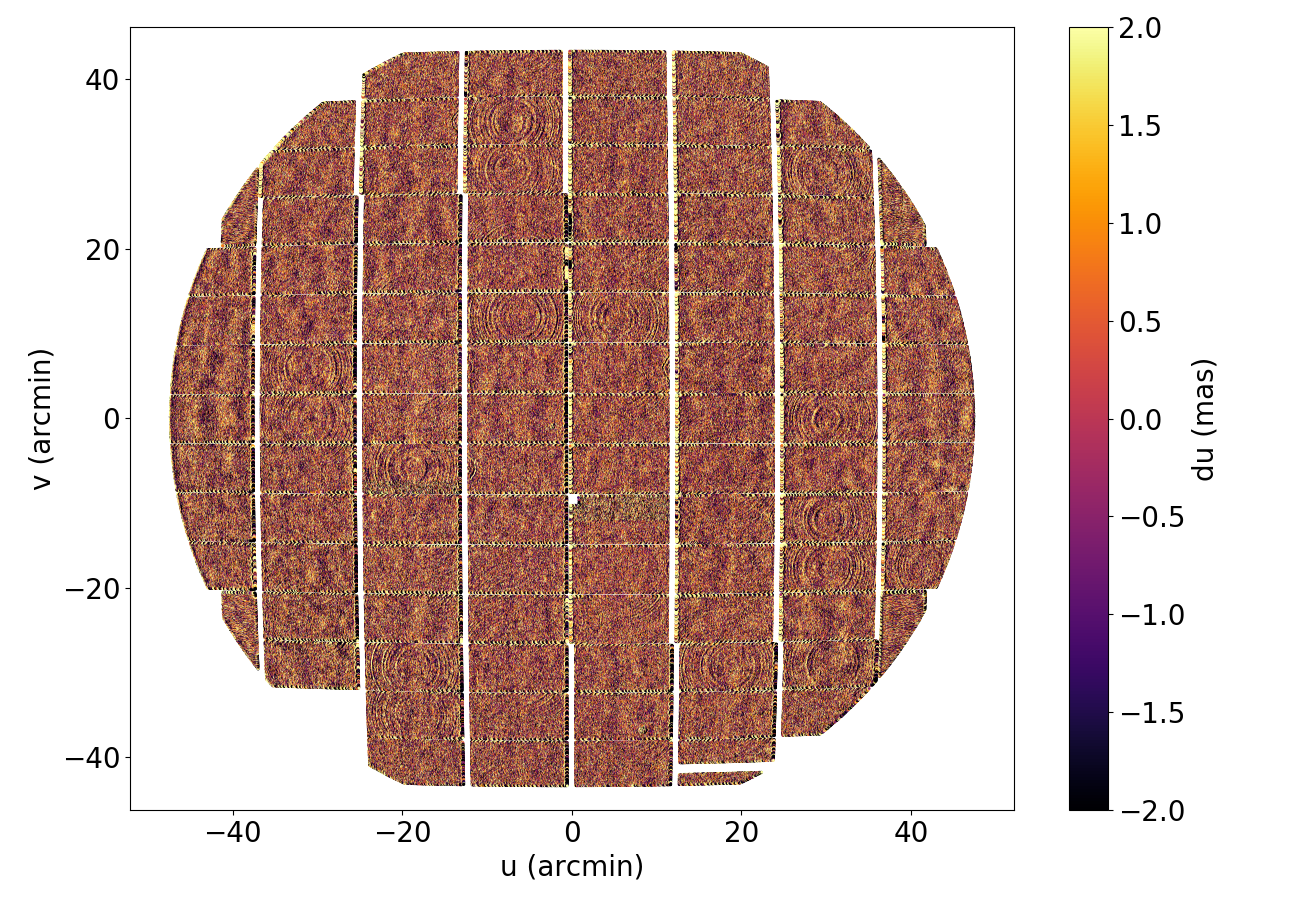}
	\includegraphics[scale=0.25]{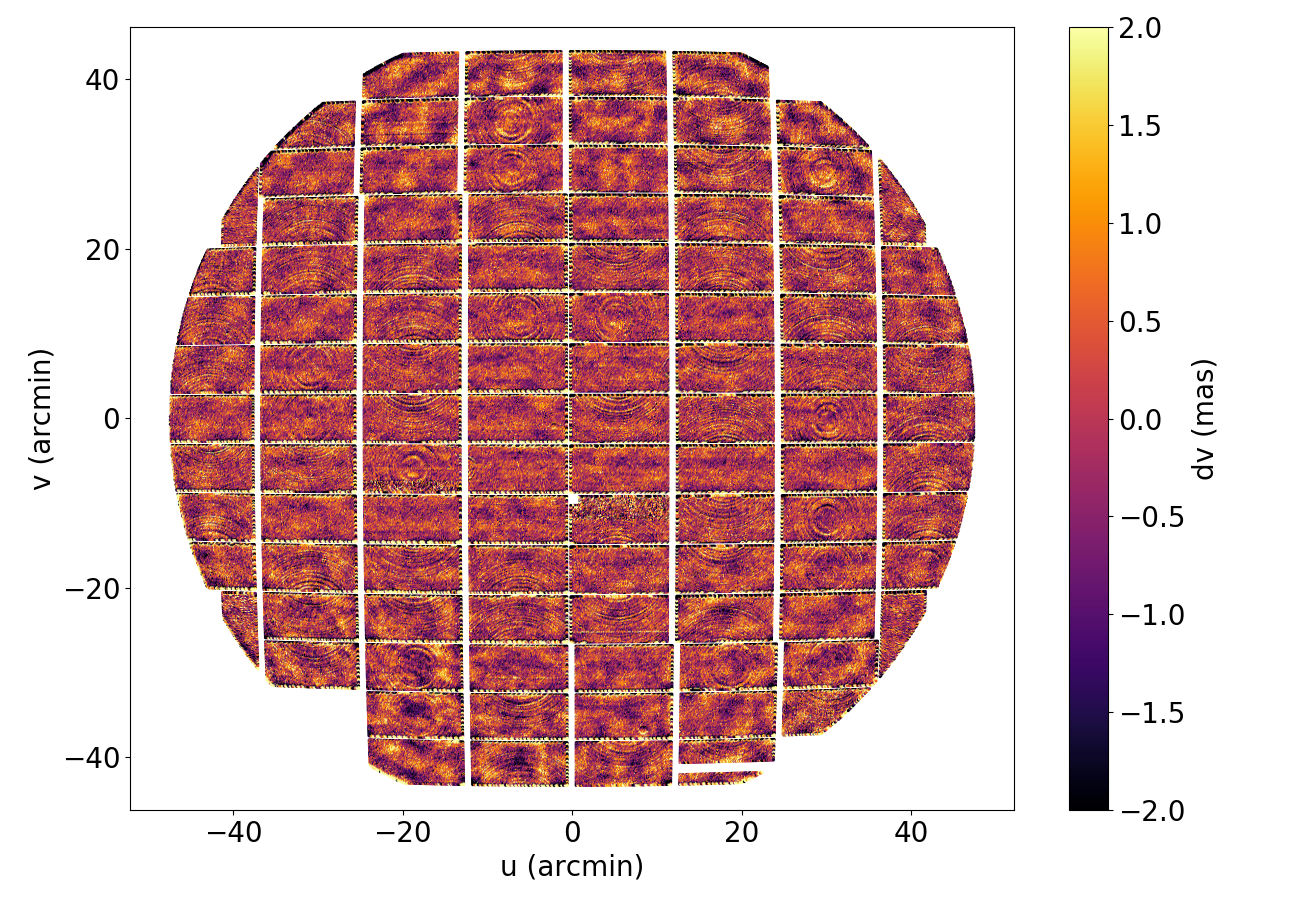}
	\includegraphics[scale=0.25]{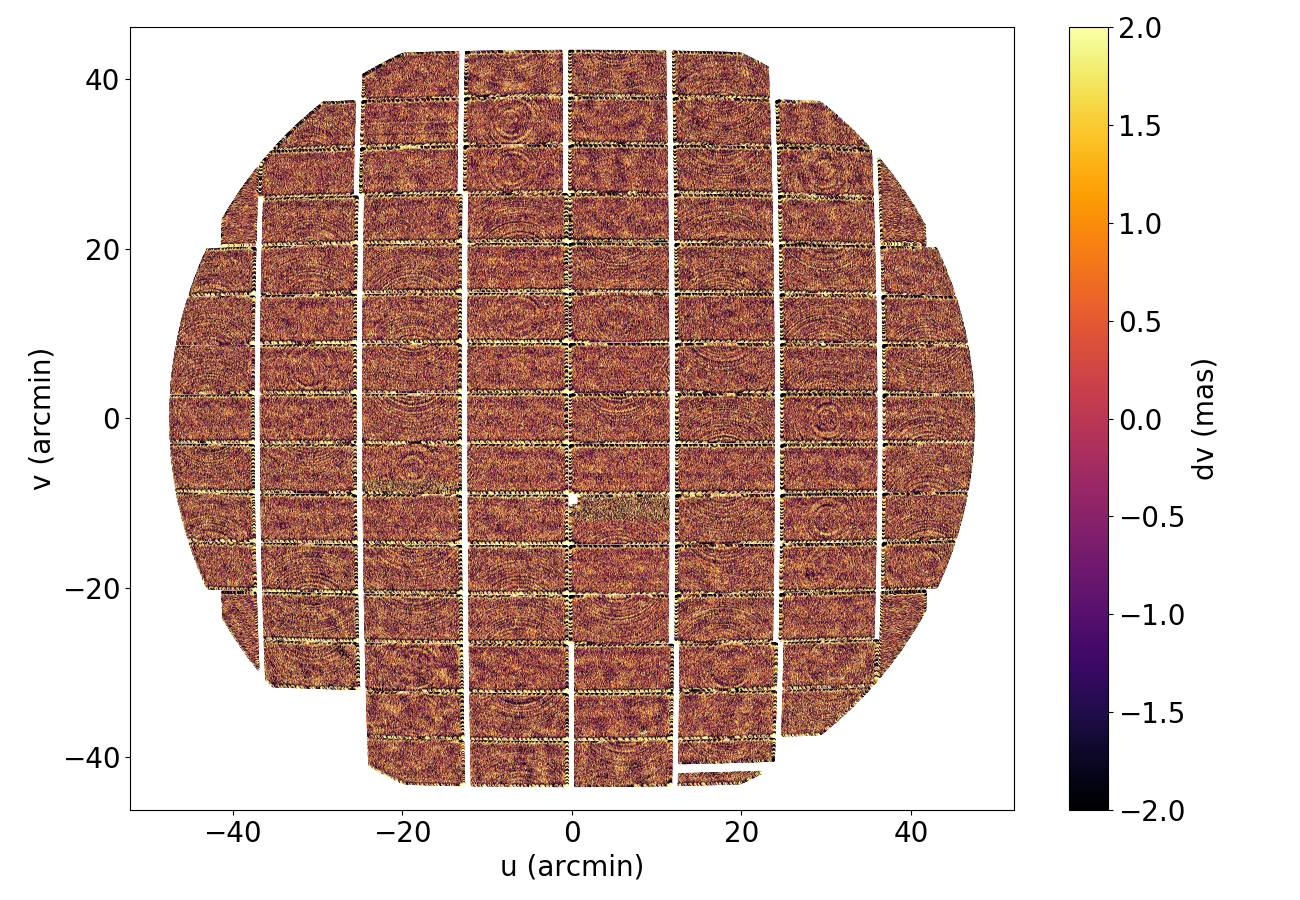}
	\caption{\small Average of the astrometric residual field for each component (top row: du, bottom row: dv) projected in the local tangent plane. 
	The per sensor transformation from pixel coordinates to tangent plane is a second-order polynomial in the left column, vs a fifth-order polynomial in the right column. }
	\label{mean_function_gp_fov}
\end{figure*}

The HSC instrument is characterized by a rapid evolution of the plate
scale in the outer part of the focal plane: the linear plate scale
varies by 10\% between the center and the edge of the field, and 70\%
of this variation is concentrated in the outer 30\% radius. In order to
verify that our modeling can cope with this variation, we plot the
residuals in the whole focal plane, in order to check for systematic
residuals at the few mas level in Fig.~\ref{mean_function_gp_fov}. 
The model consists of one polynomial transformation per sensor from pixel coordinates to the tangent plane,
common to all exposures. We see in Fig.~\ref{mean_function_gp_fov} that there are residuals when using
third-order polynomials per CCD, which
disappear with fifth-order polynomials. However, all the astrometric residuals used 
in the GP modeling used the third-order polynomials per CCD. 

\section{Impact of correlations in astrometric residuals on cosmic shear measurements}
\label{ref_disc}

\begin{figure*}
	\centering
	\includegraphics[scale=0.57]{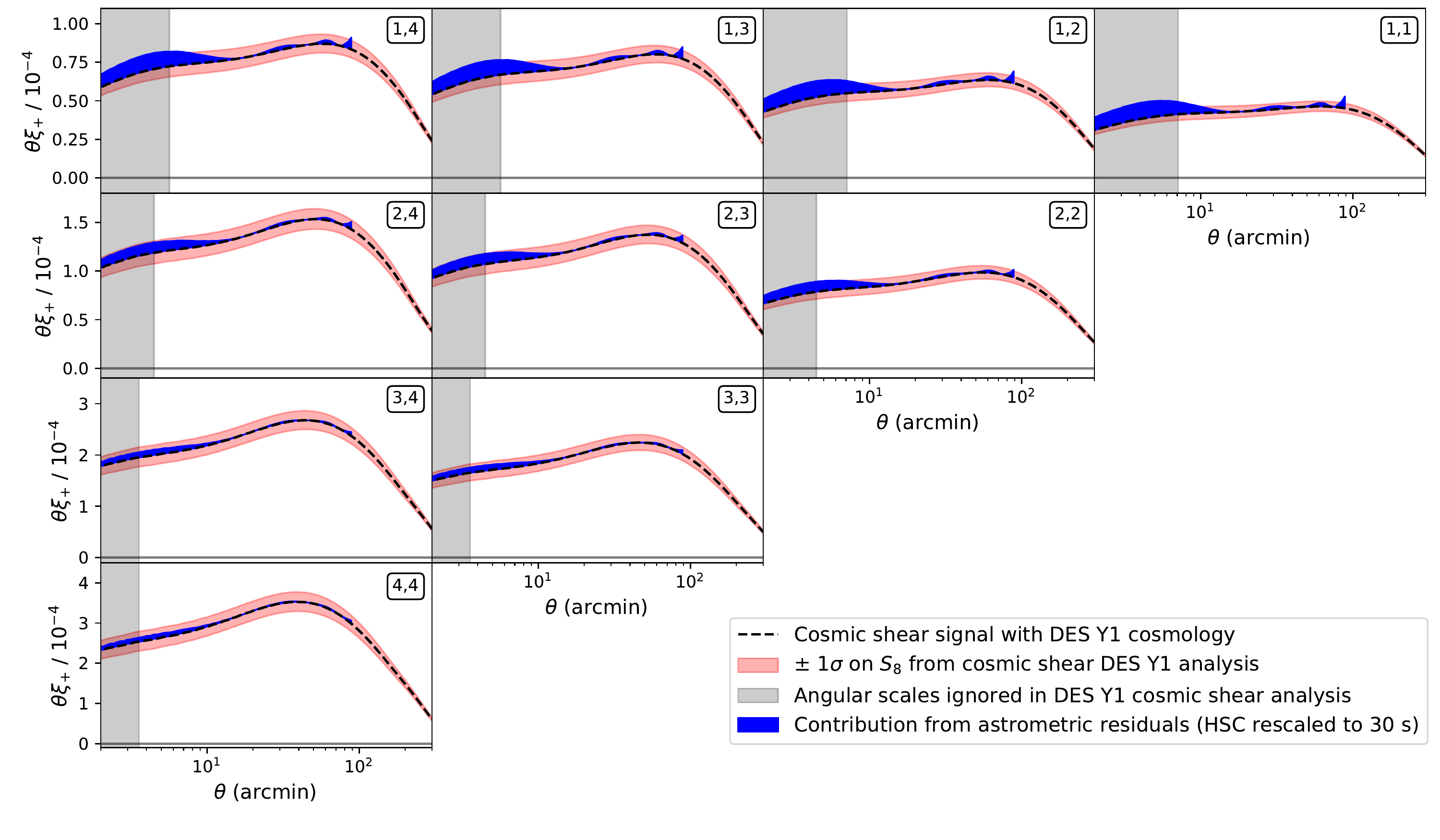}
	\caption{\small Tomographic cosmic shear signal (presented here only for $\xi_+$) 
	measured by DES Y1 (black curve) and compared to the expected signal introduced 
	by the spatial correlations of astrometric residuals due to the atmosphere by 
	rescaling HSC data on 30 s LSST exposure (blue area). If not corrected, 
	the effect of astrometric residuals in LSST on galaxy shapes will bias the 
	shear signal. This is a lower limit as the HSC site is at a higher altitude 
	that the Rubin Observatory and the effective size mirror of HSC is bigger 
	than the Rubin Observatory telescope. With the GP correction proposed in 
	this current analysis, we expect to be able to lower it by orders of magnitude.}
	\label{xipm_effects}
\end{figure*}

We explained above that Rubin Observatory will need to model the
astrometric distortions from the atmosphere because of the short
exposures (two back-to-back 15-s exposures) defined in the current
observing plan. We measure for HSC a small-scale covariance in
astrometric residuals of about 30 mas$^2$ for an average exposure
duration of 270 s. The Rubin Observatory telescope has a smaller
effective primary mirror area than the Subaru telescope and is located
at a lower elevation, so we expect atmospheric effects to be more
significant at Rubin Observatory.  Neglecting these differences, and
assuming that covariances vary only as the inverse of the exposure
time, we would expect 540 mas$^2$ and 270 mas$^2$ of small-scale
covariance on average for Rubin Observatory exposures of 15 s and 30
s, respectively. 
These covariances affect in a coherent way the
astrometric residuals, and can extend up to 1 degree, along the major
axis of the correlation function. Those may affect the measurement of
shear, and we will now provide a rough estimate of the shear
correlation purely due to these turbulence-induced residuals,
for 30-s exposures in the Rubin Observatory, assuming they are not
mitigated. 

Measuring shear without co-adding images requires 
measuring second moments using galaxy positions
averaged over images, because the position noise
biases the second moments (this is the so-called noise bias).
Using average positions is particularly
required for the Rubin observatory where the final image depth is obtained 
from hundreds of short exposures in each band.
In order to evaluate the effect of displaced positions on cosmic shear signal, 
we shift positions coherently on small spatial scales by $dx$ in a single exposure.
Using a Taylor expansion, we derive the shear offset in the  astrometric residual direction to be
\begin{equation}
\delta \gamma \simeq \frac{1}{4}\left ( \frac{dx}{\sigma_g} \right)^2,
\end{equation}
independent of the size of the PSF, where $\delta \gamma $ is the shear bias and
$\sigma_g$ denotes the r.m.s.\ angular size of the galaxy, and $dx$ the (spurious) position shift. The  general expression, accounting for actual directions, reads
\begin{equation}
\delta\gamma_1+i \delta \gamma_2 \simeq \frac{1}{4}\left ( \frac{dx+i\ dy}{\sigma_g} \right)^2.
\end{equation}
where $\gamma_1$ describes shear along the $x$ or $y$ axes, and $\gamma_2$ the shear along axes rotated by $45^\circ$, as defined in \cite{Schneider05}.
Therefore, the shear covariances induced by covariances of position offsets involve the covariance of squares of position offsets. For centered Gaussian-distributed variables $X$ and $Y$, we have
\begin{equation}
  \textrm{Cov}(X^2,Y^2) = 2 \left[ \textrm{Cov}(X,Y) \right]^2, \label{eq:cov_squared}
\end{equation}
which allows us to relate the correlation function of shear to the correlation function of position offsets.
In order to evaluate the impact of spatially correlated position offsets on
measured shear spatial correlations, we consider the smallest galaxy sizes used
in DES Y1 \citep{Zuntz18}, which have $\sigma_g \simeq 0.2''$.
Their shear will experience an  additive offset of $\sim 1.6\times 10^{-3}$ (for a small scale correlation of 270 mas$^2$), which 
will bias the shear correlation function by $\sim 6 \times 10^{-6}$, which is about half of the expected cosmic shear signal at $z\sim 0.5$ and 10~arcmin
separation.

In Fig.~\ref{xipm_effects}, we display both the
expected cosmic shear signal and the contribution from the position offsets derived from the average HSC measurement, scaled to 30-s exposures, assuming those offsets are not mitigated. 
This prediction assumes that the shear is measured on individual exposures, using a common average position with no GP modeling and correction of atmospheric displacements. The cosmic shear signal in Fig.~\ref{xipm_effects}  is computed in the four 
tomographic bins used in the cosmic shear analysis of \citealt{Troxel18}; the shear correlation functions $\xi_+$ are computed 
using the Core Cosmology Library \citep{CCL} with the fiducial cosmology result from \citealt{Troxel18} (black curve in Fig.~\ref{xipm_effects}). 
The comic shear signal presented in Fig.~\ref{xipm_effects} takes into account the non-linearity of the matter power spectrum and the spatial 
correlations introduced by the intrinsic alignment as in \citealt{Troxel18}. For a representation of the current knowledge of the cosmic shear 
signal, we indicate in Fig.~\ref{xipm_effects} with a dark region around the cosmic shear correlation function the variations 
on $S_8$ ($\sigma_8 (\Omega_m/0.3)^{0.5}$) of $\pm 0.027$, which represent the $\pm 1 \sigma$ uncertainty from the fiducial 
cosmic shear analysis of \citealt{Troxel18}. The regions shaded in red in Fig.~\ref{xipm_effects} in each tomographic bin represent the angular scales that were removed in the \citealt{Troxel18} analysis because they could be significantly biased by baryonic effects. 
The blue area in Fig.~\ref{xipm_effects} in each of the tomographic bins represent the contribution to the cosmic shear signal due to 
the average spatial correlations of the astrometric residuals due to atmospheric turbulence scaled from the HSC average measurement to 30-s exposures.
This adds to shear a contribution comparable to the DES Y1 uncertainty, which is considerably larger than the LSST precision goal. 
These correlated position offsets were not an issue for previous surveys like DES because the variance of the astrometric residuals in DES is 3 times lower than that 
expected for LSST, and therefore the contribution to the shear correlation function is 10 times lower since is scales as 
the square of the variance. It is unlikely that the wind direction
projected on the sky would average out this effect, because the wind
direction at an observatory has a preferred direction, on average.
If we are able to reduce the position
covariances down to a few mas$^2$, the effect on the shear correlation
functions becomes 2 to 3 orders of magnitude smaller than the cosmic
shear signal itself.

We have described a measurement scheme where galaxy positions are
affected by turbulence-induced offsets. If one performs an image
co-addition prior to shear measurement, the co-addition PSF will
account for the position offsets, and hence the shear measurements may be
free of the offsets described above. This is true for regions covered
by all exposures, but there are discontinuities of both PSF and the
atmospheric-induced shear field where the number of exposures involved
in the co-addition changes. Note also that the atmospheric-induced
offsets cause a PSF correlation pattern that may prove
difficult to discribe accurately. Therefore, for short exposure
times, even if one co-adds images and then measures shear, one probably has
to accurately model the spatial correlations in the atmospheric-induced 
position shifts down to arc-minute scales.

\section{Conclusions}
\label{conclusions} 
We have studied astrometric residuals for bright stars measured in exposures acquired
with the HSC instrument on the Subaru telescope, and find that these
residuals are dominated by $E$-modes.  We have developed a fast GP interpolation method to
model the  astrometric residual field induced by atmospheric turbulence.
We find that a von K\'arm\'an kernel performs better than a Gaussian
kernel, and the modeling reduces the covariances of neighboring
sources by about one order of magnitude, from 30 mas$^2$ to 3 mas$^2$ in variance,
and the variances of bright sources by about a factor of 2. Those reductions using GP interpolation are really similar 
to recent result published in \citetalias{fortino2020reducing} with the DES dataset. Based on
simulations of atmospheric distortions above the Rubin Observatory
telescope, we find that, for short exposures, the correlated
 astrometric residuals may cause a spurious contribution to shear correlations
as large as the cosmic signal. Mitigating these turbulence-induced
offsets, possibly along the lines we have sketched in this paper, will be necessary for cosmic shear analyses. 
We have shown that the spatial correlation of astrometric residuals
can be  significantly lowered by modeling and then correcting these residuals using 
GP interpolation.  We find that it is necessary to incorporate this
modeling in the astrometric fit itself (as opposed to the
 GP interpolation done in post-processing and described in Sec.~\ref{sec:model_with_gp} and Sec.~\ref{sec:results}) in order 
to achieve more precise and accurate average source positions and pixel-to-sky transformations.
The post-processing can be done along the lines sketched in the presented analysis, namely finding the hyperparameters for a chosen form of the kernel using
the two-point correlation function of residuals for each exposure and generating the  astrometric residual field from the current residuals using Eq.~\ref{gp_intep_equation}.
Our astrometric model describes the average optical distortions of the
instrument down to the mas level, and we have been able to detect the
1 to 10 mas scale CCD-induced image distortions that causes systematic
 astrometric residuals of the sources.

\begin{acknowledgements}
We recognize the significant cultural role of Mauna Kea within the indigenous Hawaiian community, and we appreciate the
opportunity to conduct observations from this revered site. The Hyper Suprime-Cam (HSC) collaboration includes the astronomical 
communities of Japan and Taiwan, and Princeton University. The HSC instrumentation and software were developed by the National 
Astronomical Observatory of Japan (NAOJ), the Kavli Institute for the Physics and Mathematics of the Universe (Kavli IPMU), the 
University of Tokyo, the High Energy Accelerator Research Organization (KEK), the Academia Sinica Institute for Astronomy and 
Astrophysics in Taiwan (ASIAA), and Princeton University. Funding was contributed by the FIRST program from the Japanese 
Cabinet Office, the Ministry of Education, Culture, Sports, Science and Technology (MEXT), the Japan Society for the Promotion of 
Science (JSPS), Japan Science and Technology Agency (JST), the Toray Science Foundation, NAOJ, Kavli IPMU, KEK, ASIAA, and 
Princeton University. This paper is based [in part] on data collected at the Subaru Telescope and retrieved from the HSC data 
archive system, which is operated by Subaru Telescope and Astronomy Data Center (ADC) at National Astronomical Observatory of 
Japan. Data analysis was in part carried out with the cooperation of Center for Computational Astrophysics (CfCA), National Astronomical 
Observatory of Japan. We gratefully acknowledge support from the CNRS/IN2P3 Computing Center (Lyon - France) for providing computing 
and data-processing resources needed for this work. We acknowledge useful discussions with Gary Bernstein, Nao Suzuki, and Naoki Yasuda. 
We thank Patricia Burchat, Josh Meyers, and Claire-Alice H\'ebert for reviewing and giving helpful advice on this paper. PFL acknowledges support from the 
National Science Foundation grant PHY-1404070.
\end{acknowledgements}

\bibliographystyle{aa}
\bibliography{Biblio}

\nopagebreak

\appendix

\section{Integral of the correlation function}
\label{integral_correlation_function}

We  demonstrate here that the integral of the correlation function for astrometric residuals  
is null because the average of the astrometric residual field is null per construction of the model. In the continuous limit, the correlation function is defined as 
\begin{equation}
\xi(r) = \int f(x-r) f(x) dx .
\end{equation}
Consequently, the integral of the correlation function is
\begin{equation}
\int \xi(r) dr = \int \int f(x-r) f(x) dx dr .
\end{equation}
Making the change of variables $X = x-r$, we have
\begin{equation}
\int \xi(r) dr = \left( \int f(X) dX \right)\left( \int f(x) dx \right) .
\end{equation}
For the specific case of astrometric residuals, the average of the astrometric residual field is constructed to be zero; 
consequently, we have
\begin{equation}
\int f(x) dx = 0,
\end{equation}
which means the integral over the correlation function is zero:
\begin{equation}
\int \xi(r) dr = 0.
\end{equation}

\end{document}